\documentclass[12pt, leqno]{article}
\usepackage{amsmath,setspace,multirow,lineno}
\usepackage[font=small,labelfont=bf,singlelinecheck=off]{caption}
\usepackage[top=1.5in, bottom=1.5in, left=1.1in, right=1.1in]{geometry}

\usepackage[round]{natbib}
\usepackage{color,soul}

\DeclareCaptionStyle{italic}[justification=centering]{labelfont={bf},textfont={it},labelsep=colon}

\usepackage{graphicx,psfrag,epsf,textcomp,epstopdf, amsthm, paralist}
\usepackage{enumerate, dsfont, alltt, verbatim}
\usepackage{natbib}
\usepackage{url,xcolor}
\usepackage{booktabs, subfig, bm, paralist,mathpazo,tikz,longtable,microtype, authblk}
\usepackage[linesnumbered,ruled,vlined]{algorithm2e}

\usepackage[pdftex,colorlinks=true]{hyperref}
\definecolor{darkblue}{rgb}{0,0,.6}
\hypersetup{citecolor=darkblue,linkcolor=darkblue,urlcolor=darkblue}
\definecolor{DarkRed}{rgb}{.7,0,.4}

\usepackage{comment,orcidlink}

\newcommand{\blind}{0}
\DeclareMathOperator{\atantwo}{atan2}

\newcommand{\X}{\mathcal{X}}
\newcommand{\Y}{\mathcal{Y}}

\newcommand{\Rlogo}{\protect\includegraphics[height=1.8ex,keepaspectratio]{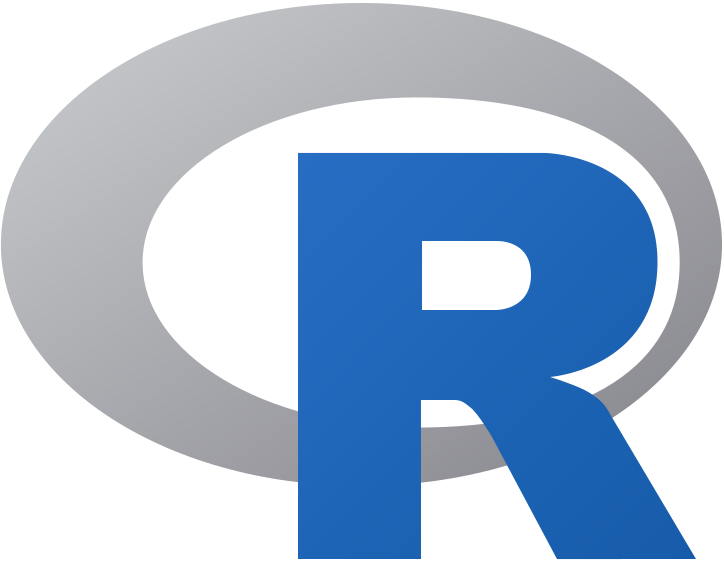}}

\graphicspath{{plots/}}
\DeclareMathOperator*{\argmin}{\arg\!\min}

\newsavebox\CBox

\newtheorem{@definition}{\sc Definition}[section]

\newtheorem{proposition}{\sc Proposition}[section]

\newtheorem{theorem}{\sc Theorem}[section]
\newtheorem{remark}{\sc Remark}[section]

\renewcommand\X{\mathcal{X}}

\date{}

\begin{document}

\def\spacingset#1{\renewcommand{\baselinestretch}{#1}\small\normalsize} \spacingset{1}

\if0\blind
{
\title{\bf Spatial function-on-function regression}}
\author[1]{\normalsize Ufuk Beyaztas\thanks{Corresponding address: Department of Statistics, Marmara University, 34722, Kadikoy-Istanbul, Turkey; Email: ufuk.beyaztas@marmara.edu.tr} \orcidlink{0000-0002-5208-4950}}
\author[2]{\normalsize Han Lin Shang \orcidlink{0000-0003-1769-6430}}
\author[1]{\normalsize Gizel Bakicierler Sezer \orcidlink{0000-0002-1789-0842}}
\author[3]{\normalsize Abhijit Mandal \orcidlink{0000-0003-3830-6526}}
\author[4]{\normalsize Roger S. Zoh \orcidlink{0000-0002-8066-1153}}
\author[4]{\normalsize Carmen D. Tekwe \orcidlink{0000-0002-1857-2416}}

\affil[1]{\normalsize Department of Statistics, Marmara University, Turkey}
\affil[2]{\normalsize Department of Actuarial Studies and Business Analytics, Macquarie University, Australia}
\affil[3]{\normalsize Department of Mathematical Sciences, University of Texas at El Paso, USA}
\affil[4]{\normalsize Department of Epidemiology and Biostatistics, Indiana University, School of Public Health, USA}
\maketitle
\fi

\if1\blind
{
\title{\bf Spatial function-on-function regression}
\author{}
} \fi

\maketitle

\begin{abstract}
We introduce a spatial function-on-function regression model to capture spatial dependencies in functional data by integrating spatial autoregressive techniques with functional principal component analysis. The proposed model addresses a critical gap in functional regression by enabling the analysis of functional responses influenced by spatially correlated functional predictors, a common scenario in fields such as environmental sciences, epidemiology, and socio-economic studies. The model employs a spatial functional principal component decomposition on the response and a classical functional principal component decomposition on the predictor, transforming the functional data into a finite-dimensional multivariate spatial autoregressive framework. This transformation allows efficient estimation and robust handling of spatial dependencies through least squares methods. In a series of extensive simulations, the proposed model consistently demonstrated superior performance in estimating both spatial autocorrelation and regression coefficient functions compared to some favorably existing traditional approaches, particularly under moderate to strong spatial effects. Application of the proposed model to Brazilian COVID-19 data further underscored its practical utility, revealing critical spatial patterns in confirmed cases and death rates that align with known geographic and social interactions. The \texttt{[package anonymized for review]} package in \Rlogo{} \ provides a comprehensive implementation of the proposed estimation method, offering a user-friendly and efficient tool for researchers and practitioners to apply the methodology in real-world scenarios.
\end{abstract}


\newpage
\spacingset{1.65} 

\section{Introduction} \label{sec:1}

The concept of a functional linear regression model, which accommodates functional responses alongside functional and/or scalar predictors, was first introduced by \cite{ramsay1991}. Since then, a wide range of methodologies for parameter estimation in this model have been developed, demonstrating their effectiveness across various disciplines, as evidenced by studies such as \cite{yao2005}, \cite{MullerYao2008}, \cite{wang2014}, \cite{ivanescu2015}, \cite{BS20}, \cite{BS20C}, \cite{Cai2022}, and \cite{wang2022}. This research is centered on a function-on-function regression (FoFR) framework tailored to model functional outcomes by incorporating functional covariates.

Consider a random sample $(\Y_i, \X_i)$, $i \in \{1, \ldots, n \}$ drawn from the joint distribution of $(\Y, \X)$, where $\Y = \Y(t)$, $t \in \mathcal{I}_y$, represents the functional response and $\X = \X(s)$, $s \in \mathcal{I}_x$, denotes the functional covariate. In this context, the functional response and the functional covariate are considered stochastic processes over the closed and bounded intervals $t \in \mathcal{I}_y$ and $s \in \mathcal{I}_x$, respectively, meaning $\Y_i(t): \mathcal{I}_y \rightarrow \mathbb{R}$ and $\X_i(s): \mathcal{I}_x \rightarrow \mathbb{R}$ for each $i \in \{1, \ldots, n \}$. The FoFR model considered can be expressed as:
\begin{equation}\label{eq:ffrm}
\Y_i(t) = \beta_0(t) + \int_{\mathcal{I}_x} \X_i(s) \beta(s,t) ds + \epsilon_i(t),
\end{equation}
where $\beta_0(t): \mathcal{I}_y \rightarrow \mathbb{R}$ is the intercept function, $\beta(s,t): \mathcal{I}_x \times \mathcal{I}_y \rightarrow \mathbb{R}$ is the bivariate regression coefficient function, and $\epsilon_i(t)$ is the random error term, assumed to satisfy $\mathbb{E}\{\epsilon(t)\} = 0$ and $\mathrm{Var} \{\epsilon(t)\} < \infty$ for all $t \in \mathcal{I}_y$.

The FoFR model in~\eqref{eq:ffrm} has been adapted in various forms, such as function-on-function and function-on-scalar models, and extended to nonlinear frameworks \citep[see, e.g.,][]{Lian2007, Schepl2015, Kim2018, BSQuad, rao2023, BSM2024}, as well as nonparametric approaches \citep[see, e.g.,][]{wang2019, wang2022JCG, Boumahdi2023}. A common assumption in these models is the independence of individual data elements. However, this assumption often does not hold in fields such as population health, demography, economics, environmental sciences, agronomy, and mining, where datasets typically exhibit spatial correlation.

Numerous techniques have been devised to examine data from discrete matrices exhibiting spatial dependence \citep[see, e.g.][]{Ord1975, Anselin1998, Lesage2009, Cressie2015, Schabenberger2017, Wang2019c}. Additionally, there has been significant research into spatially correlated functional data \citep[see, e.g.,][]{Nerini2010, Giraldo2011, Zhang2011, Caballero2013, Zhang2016, Bohorquez2016, Menafoglio2017, Aguilera2017, Bohorquez2017, Giraldo2018}. The majority of these studies employ kriging methods tailored to point-referenced data. This paper narrows its scope to exploring the FoFR model within the context of areal data.

In analyzing spatial dependence in areal data, three key spatial linear regression models are often employed: the spatial autoregressive model, the spatial error model, and the spatial Durbin model \citep{Lesage2009}. The spatial autoregressive model is particularly noteworthy for its straightforward interpretation, as it involves a single parameter representing spatial dependence through the spatial lag of the response variable \citep{Huang2021}. This characteristic simplifies extending developments from this model to the other two. Given these advantages, our research focuses on the spatial autoregressive model, specifically extending it to the FoFR model to effectively handle spatial dependencies in functional areal data.

Only a few methods have been developed to address spatial dependence in functional regression models involving a functional response variable. \cite{Zhu2022} introduced a social network model within a time series framework, where the response variable varies over time, assuming that interactions occur only at concurrent time points, without influence from past or future values. On the other hand, \cite{Hoshino2024} presented a spatial autoregressive model that treats the dependent variable as a function but incorporates only scalar covariates, focusing on estimating the conditional quantile of the functional response based on these covariates. However, no method has been proposed to estimate the spatial lag parameter and the regression coefficient functions in the FoFR framework presented in~\eqref{eq:ffrm}.

This research addresses a significant gap by presenting an innovative spatial FoFR (SFoFR) model. The proposed model treats the response variable as a stochastic process with complex temporal dynamics and accommodates a functional predictor. To estimate the model parameters, we first perform a spatial functional principal component (SFPC) analysis \citep{Khoo2023} on the functional response variable and a classical functional principal component (FPC) analysis on the functional predictor variable. These decompositions enable the representation of functional elements within a finite-dimensional space defined by the SFPC and FPC coefficients, thereby transforming the infinite-dimensional problem into a finite-dimensional one. Using these coefficients, the SFoFR model is converted into a multivariate spatial autoregressive model. We then apply the least-squares estimator developed by \cite{Zhu2020}, which provides computational efficiency and robust parameter estimation for the resulting multivariate spatial autoregressive model. This approach ensures that complex functional dependencies and spatial correlations are effectively captured and modeled. The $\sqrt{n}$-consistency and asymptotic normality of the proposed estimator are derived under some regulatory conditions.

The structure of the paper is as follows: Section~\ref{sec:2} introduces the proposed SFoFR model, discusses its completeness property, and defines the associated model parameters. Section~\ref{sec:3} provides a detailed description of the estimation procedure for the model parameters. In Section~\ref{sec:4}, the estimation and predictive performance of the proposed method are evaluated through Monte Carlo simulations, with results presented and described. Section~\ref{sec:5} applies the proposed method to an empirical data example, with a discussion of the findings. Finally, Section~\ref{sec:6} provides some concluding remarks and future directions. Additional technical details regarding the model's completeness property and the asymptotic normality of the estimators are provided in the online supplementary material.

\section{Model, notations, and nomenclature}\label{sec:2}

We begin by introducing some notation that will be utilized throughout this manuscript. Let $\mathbb{I}_n$ represent the $n \times n$ identity matrix. For any function $x(t)$ defined on the closed and bounded interval $t \in \mathcal{I}$, the $\mathcal{L}^p$ norm of $x(t)$ is given by $\Vert \phi(t) \Vert_{\mathcal{L}^p}:= \{\int_{\mathcal{I}} \vert x(t) \vert^p dt \}^{1/p}$, and $\mathcal{L}^p(\mathcal{I})$ denotes the set of functions $x(t)$ for which $\Vert x(t) \Vert_{\mathcal{L}^p} < \infty$. For a matrix $\bm{M} \in \mathbb{R}^{n \times n}$, $\lambda_i(\bm{M})$ represents the $i^\textsuperscript{th}$ eigenvalue of $\bm{M}$, ordered such that $\vert \lambda_1(\bm{M})\vert \geq \ldots \geq \vert \lambda_n(\bm{M})\vert$. Furthermore, $\Vert \bm{M} \Vert_F$ and $\Vert \bm{M} \Vert_{\infty}$ refer to the Frobenius norm and the maximum absolute row sum of $\bm{M}$, respectively. For any vector $v$, let $v_{-(j)}$ denote the vector excluding $j^\textsuperscript{th}$ element.

Consider a stochastic process $\{Y_v(t), \X_v(s)\}$, $v \in \mathcal{D} \subset \mathbb{R}^q$ and $q \geq 1$, observed from discrete, evenly or unevenly spaced lattice subsets $\mathcal{D}$, consisting of spatial units $v_1, \ldots, v_n$. Here, for some $2 \leq p < \infty$, $Y_v(t) \in \mathcal{L}^p(\mathcal{I}_y)$ and $\X_v(s) \in \mathcal{L}^p(\mathcal{I}_x)$ for all $v$. For simplicity, we use the notation $i$ to refer to spatial unit $v_i$. Additionally, we assume $\mathcal{I}_y = \mathcal{I}_x = [0,1]$, so $Y_v(t)$ and $\X_v(s)$ are defined on $[0,1]$ and map to $\mathbb{R}$. The i.i.d. random samples are mean-zero stochastic processes, meaning $\mathbb{E}\{\Y(t)\} = \mathbb{E}\{\X(s)\} = 0$. We then consider the following form for the SFoFR model:
\begin{equation}\label{eq:sfofrm}
\Y_i(t) = \sum_{i^{\prime}=1}^n w_{i i^{\prime}} \int_0^1 \Y_{i^{\prime}}(u) \rho(u,t) du + \int_0^1 \X_i(s) \beta(s,t) ds + \epsilon_i(t),
\end{equation}
where $w_{i i^{\prime}} \in \mathbb{R}^{+}$, $i, i^{\prime} \in \{1, \ldots, n \}$ represents the $(i i^{\prime})^\textsuperscript{th}$ element of an $n \times n$ pre-specified spatial weight matrix $\bm{W} = (w_{i i^{\prime}})_{n \times n}$. Each spatial weight $w_{i i^{\prime}}$ denotes the spatial relationship between locations $i$ and $i^{\prime}$ ($w_{ii} = 0$). The function $\rho(u,t) \in \mathcal{C}[0,1]^2$, where $\mathcal{C}[0,1]^2$ is the set of continuous functions on $[0,1]^2$, represents the unknown spatial autocorrelation parameter function. The function $\beta(s,t) \in \mathcal{C}[0,1]^2$ signifies the bivariate regression coefficient function, and $\epsilon_i(t)$ is the random functional error term.

In Model~\eqref{eq:sfofrm}, the spatial autocorrelation parameter function $\rho(u,t)$ plays a crucial role in measuring the degree of spatial dependence among neighboring locations. This model hinges on the definition of spatial weights $w_{i i^{\prime}} \in \mathbb{R}$, $i, i^{\prime} \in \{1, \ldots, n \}$. The arrangement of these $n$ spatial units can vary widely, spanning regular or irregular distributions across any spatial domain, provided the construction of the matrix $\bm{W}$ remains viable. When data are gathered from a regular grid, neighboring units typically share borders, corners, or both \citep{Anselin1998}. Conversely, with irregular grids, units are considered neighbors if they share common edges \citep{Huang2021}.

The entries of the symmetric weight matrix $\bm{W}$, denoted as $w_{i i^{\prime}}$, can be formulated using a distance metric, such that $w_{i i^{\prime}} = m(d_{i i^{\prime}})$, where $m(\cdot)$ is a monotonically decreasing function, and $d_{i i^{\prime}}$ represents the distance between units $i$ and $i^{\prime}$ ($d_{ii} = 0$). This distance metric could encapsulate geographical, economic, social, or policy-related factors, or a blend thereof \citep[see, e.g.,][]{Yu2016}. Importantly, the spatial weight matrix does not necessarily need to be symmetric; asymmetrical weight matrices can also be relevant depending on the specific characteristics of the data, as discussed in \citep{Huang2021}. Typically, $\bm{W}$ is normalized row-wise, ensuring $w_{i i^{\prime}} = \frac{ m(d_{i i^{\prime}})}{\sum_{j^{\prime}=1}^n m(d_{i j^{\prime}})}$, thereby guaranteeing that each row sums to unity, with zero values on the diagonal.

Let $\Y(t) = [\Y_1(t), \ldots, \Y_n(t)]^\top$, $\X(s) = [\X_1(s), \ldots, \X_n(s)]^\top$, and $\epsilon(t) = [\epsilon_1(t), \ldots, \epsilon_n(t)]^\top$. Consequently, Model~\eqref{eq:sfofrm} can be reformulated in matrix form as:
\begin{equation}\label{eq:Mform}
\Y(t) = \bm{W} \int_0^1 \Y(u) \rho(u,t) du + \int_0^1 \X(s) \beta(s,t) ds + \epsilon(t).
\end{equation}
Before delving into parameter estimation for Model~\eqref{eq:Mform}, we explore its structural completeness with the following proposition. 

\begin{proposition}\label{prop1}
Let $\mathcal{T}: (\mathcal{L}^p)^n[0,1] \rightarrow (\mathcal{L}^{p})^n[0,1]$ be the linear operator defined by
\begin{equation*}
(\mathcal{T} \Y)(t):= \bm{W} \int_0^1 \Y(u) \rho(u,t) du,
\end{equation*}
where $(\mathcal{L}^p)^n[0,1]$ is the space of vector-valued functions $\Y(t) = [\Y_1(t), \ldots, \Y_n(t)]^\top$ such that each component $\Y_i(t)$ is in $\mathcal{L}^p[0,1]$ and the norm on $(\mathcal{L}^p)^n[0,1]$ is defined as $\Vert \Y \Vert_{(\mathcal{L}^p)^n} := \left( \sum_{i=1}^n \Vert \Y \Vert_{\mathcal{L}^p}^n \right)^{1/p}$. Define the identity operator $\mathbb{I}_d := (\mathcal{L}^p)^n[0,1] \rightarrow (\mathcal{L}^p)^n[0,1]$ by $(\mathbb{I}_d \Y)(t) = \Y(t)$.

Suppose $\Vert \rho \Vert_{\infty} < \frac{1}{\Vert \bm{W} \Vert_{\infty}}$, where $\Vert \rho \Vert_{\infty}:= \max_{(u,t) \in [0,1]^2} \vert \rho(u,t) \vert$. Then, the operator $\mathbb{I}_d - \mathcal{T}$ is invertible, and $(\mathbb{I}_d - \mathcal{T})^{-1}$ exists. Consequently, the solution $\Y(t)$ to the system in~\eqref{eq:Mform} is unique and can be expressed as
\begin{equation*}
\Y(t) = (\mathbb{I}_d - \mathcal{T})^{-1} \left\lbrace \int_0^1 \X(s) \beta(s,t) ds + \epsilon(t) \right\rbrace.
\end{equation*}
\end{proposition}
The proof of Proposition~\ref{prop1} is deferred to the online supplementary material.

\section{Estimation}\label{sec:3}

Let us explore the matrix representation of our proposed framework, as defined by~\eqref{eq:Mform}. Within this framework, the direct estimation of the regression coefficient functions $\rho(u,t)$ and $\beta(s,t)$ presents a challenging inverse problem. Therefore, in our estimation procedure, we start by projecting all functional entities from~\eqref{eq:Mform} onto a finite-dimensional space using the FPC-based dimension reduction technique. This approach effectively converts the infinite-dimensional structure described in~\eqref{eq:Mform} into a finite-dimensional model of multivariate spatial autocorrelation, utilizing basis expansion coefficients.

Consider an arbitrary (centered) function $x(t) = [x_1(t), \ldots, x_n(t)]^\top$ defined on the interval $[0,1]$. The covariance function of $x(t)$ is denoted by $G(s,t) = \mathrm{Cov}[x(s), x(t)]$. According to the Karhunen-Lo\'eve theorem, the covariance kernel $G(s,t)$ has an eigen-decomposition as follows:
\begin{equation*}
G(s,t) = \sum_{k=1}^{\infty} \delta_k \eta_k(s) \eta_k(t), \quad t,s \in [0,1],
\end{equation*}
where $\eta_k(t)$, $k \in \{1, 2, \ldots \}$, are orthonormal eigenfunctions (also referred to as FPCs) corresponding to non-negative eigenvalues $\delta_k$, $k \in \{1, 2, \ldots \}$, with $\delta_k \geq \delta_{k+1}$. It is assumed that the eigenvalues of $G(s,t)$ are distinct to ensure the uniqueness of orthonormal bases of eigenfunctions. In practice, most variability in functional variables can be captured by a finite number of the first few eigenfunctions. Thus, we adopt a strategy of projecting $x(t)$ onto basis expansions with a predefined truncation constant $K$. Consequently, each realization $x_i(t)$ ($i \in \{1, \ldots, n\}$) can be represented by an expansion $x(t) \approx \sum_{k=1}^K \xi_{ik} \eta_k(t)$, where the uncorrelated random variables $\xi_{ik} = \int_0^1 x_i(t) \eta_k(t) dt$ (also known as FPC scores) denote the projections of $x(t)$ onto their respective orthonormal bases.

In practical applications, even though $x(t)$ resides in an infinite-dimensional space, it is typically observed at discrete time points. Functional forms of discretely observed functions are often approximated using basis expansion methods, such as B-splines or Fourier bases. Here, we focus on the B-spline basis expansion because of its simplicity and effectiveness. Let $x_i(t) = \sum_{l=1}^{L_x} a_{il} \gamma_l(t)$ denote the B-spline basis expansion for $x_i(t)$, where $\gamma_l(t)$, $l \in \{1, \ldots, L_x \}$, represents the B-spline basis functions and $a_{il}$, $l \in \{1, \ldots, L_x \}$ and $i \in \{1, \ldots, n \}$, are the corresponding expansion coefficients. For $x(t) = [x_1(t), \ldots, x_n(t)]^\top$, we can express $x(t) = \bm{A} \gamma(t)$, where $\bm{A} = (a_{il})_{il} \in \mathbb{R}^{n \times L_x}$ and $\gamma(t) = [\gamma_1(t), \ldots, \gamma_{L_x}(t)]^\top$. Consequently, the covariance function is given by $\widehat{G}(s,t) = n^{-1} x^\top(s) x^\top(t) = n^{-1} \gamma^\top(s) \bm{A}^\top \bm{A} \gamma(t)$. Assuming $\eta(t) = [\eta_1(t), \ldots, \eta_K(t)]^\top$ can be expanded in terms of B-splines as $\eta(t) = \sum_{l=1}^{L_x} \eta_l \gamma_l(t)$ with basis expansion coefficients $\eta_l$, $l \in \{1, \ldots, L_x\}$, the FPC analysis in the basis expansion framework is expressed as follows:
\begin{align*}
\int_0^1 \widehat{G}(s,t) \eta(t) &= \int_0^1 n^{-1} \gamma^\top(s) \bm{A}^\top \bm{A} \gamma(t) \eta(t) dt = n^{-1} \gamma^\top(s) \bm{A}^\top \bm{A} \int_0^1 \gamma(t) \eta(t) dt, \\
&= n^{-1} \gamma^\top(s) \bm{A}^\top \bm{A} \sum_{l=1}^{L_x} \eta_k \int_0^1 \gamma(t) \gamma_l(t) dt = n^{-1} \gamma^\top(s) \bm{A}^\top \bm{A} \bm{\Gamma} \bm{\eta},
\end{align*}
where $\bm{\Gamma} = \int_0^1 \gamma(t) \gamma^\top(t) dt \in \mathbb{R}^{L_x \times L_x}$ and $\bm{\eta} = [\eta_1, \ldots, \eta_{L_x}]^\top$. 

The aforementioned results demonstrate that solving the eigenanalysis in an infinite-dimensional space can be reduced to solving it in a finite-dimensional space of coefficients obtained from basis expansion. Specifically, $n^{-1} \gamma^\top(s) \bm{A}^\top \bm{A} \Gamma \bm{\eta} = \delta \gamma^\top(s) \bm{\eta}$ simplifies to $n^{-1} \bm{A}^\top \bm{A} \bm{\Gamma} \bm{\eta} = \delta \bm{\eta}$. Introducing $\bm{u} = \bm{\Gamma}^{1/2} \bm{\eta}$, the eigenanalysis decomposes as follows:
\begin{align*}
n^{-1} \bm{\Gamma}^{1/2} \bm{A}^\top \bm{A} \bm{\Gamma}^{1/2} \bm{\Gamma}^{1/2} \bm{\eta} &= \delta \bm{\Gamma}^{1/2} \bm{\eta}, \\
n^{-1} \bm{\Gamma}^{1/2} \bm{A}^\top \bm{A} \bm{\Gamma}^{1/2} \bm{u} &= \delta.
\end{align*}
In essence, each FPC corresponds to performing multivariate principal component analysis on the matrix $\bm{D} = \bm{\Gamma}^{1/2} \bm{A}$. Let $\bm{\chi} = [\chi_1, \ldots, \chi_{L_x}]^{\top}$ denote the matrix of orthogonal eigenvectors that maximize the variances of the scores $\bm{\kappa} = \bm{D} \bm{\chi}$:
\begin{equation*}
\mathrm{Var}(\bm{\kappa}) = n^{-1} \bm{\chi}^\top \bm{D}^\top \bm{D} \bm{\chi}.
\end{equation*}
Consequently, the FPCs are expressed as $\widehat{\eta}(t) = \bm{\chi} \gamma(t)$, and their corresponding scores $\bm{\xi} = [\xi_1, \ldots, \xi_K]^{\top}$ are obtained as $\bm{\xi} = \int_0^1 x(t) \widehat{\eta}(t) dt$.

In traditional FPC analysis, the assumption is that the functions $x_i(t)$ are independent across different indices $i$, leading to $\text{Cor}(\xi_i, \xi_j) = 0$ for distinct curves. However, this assumption becomes impractical when spatial dependencies exist among the realizations of $x(t)$. Therefore, for the functional response, we turn to the SFPC analysis proposed by \cite{Khoo2023}. 

SFPC analysis aims to derive components and their corresponding scores that effectively capture both variability and spatial structures within the data \citep{Khoo2023}. In SFPC, the goal is to identify orthogonal eigenvectors $\widetilde{\bm{\chi}} = [\widetilde{\chi}_1, \ldots, \widetilde{\chi}_{L_x}]^\top$, such that the scores $\widetilde{\bm{\kappa}} = \bm{D} \widetilde{\bm{\chi}}$ exhibit both dispersion and spatial autocorrelation. This approach seeks to maximize the following criterion:
\begin{equation}\label{eq:vspca}
\mathrm{Var}(\widetilde{\bm{\kappa}}) I(\widetilde{\bm{\kappa}}) = n^{-1} \widetilde{\bm{\chi}}^\top \bm{D}^\top \bm{W} \bm{D} \widetilde{\bm{\chi}},
\end{equation}
where $I(\widetilde{\bm{\kappa}}) = \frac{\widetilde{\bm{\kappa}}^\top \bm{W} \widetilde{\bm{\kappa}}}{\widetilde{\bm{\kappa}}^\top \widetilde{\bm{\kappa}}}$ represents the Moran's I statistic \citep{Eckardt2021}, and $\bm{W}$ denotes the spatial weight matrix. The SFPCs are obtained as $\widetilde{\widehat{\eta}}(t) = \widetilde{\bm{\chi}} \gamma(t)$ and their respective scores are given by $\widetilde{\bm{\xi}} = [\widetilde{\xi}_1, \ldots, \widetilde{\xi}_K]^\top = \int_0^1 x(t) \widetilde{\widehat{\eta}}(t) dt$. The components of $\widetilde{\bm{\chi}}$ derived from the criterion in~\eqref{eq:vspca} are related to $n^{-1} \bm{D}^\top \bm{W} \bm{D}$, which differs from classical principal components due to the involvement of $\bm{W}$ and its impact on positive definiteness \citep[see, e.g.,][for further discussion]{Jombart2008}. Consequently, this may result in some principal component scores associated with negative eigenvalues. The criterion in~\eqref{eq:vspca} exhibits high positivity when $\widetilde{\bm{\kappa}}$ shows significant variance and a broad spatial structure, while it tends to be highly negative when $\widetilde{\bm{\kappa}}$ has substantial variance but localized spatial characteristics.

\subsection{FPC decomposition of infinite-dimensional SFoFR model}

To start with, let us consider the FPC and SFPC decompositions of all the functional objects in~\eqref{eq:Mform} as follows:
\begin{align}
\Y_i(t) &\approx \sum_{k_1=1}^{K_y} y_{i k_1} \phi_{k_1}(t) = \bm{y}_i^\top \phi(t), \quad \forall t \in [0,1], \label{eq:app1} \\
\X_i(s) &\approx \sum_{k_2=1}^{K_x} x_{i k_2} \psi_{k_2}(s) = \bm{x}_i^\top \psi(s), \quad \forall s \in [0,1], \label{eq:app2} \\
\rho(u,t) &\approx \sum_{k_1^{\prime}=1}^{K_y} \sum_{k_1=1}^{K_y} \rho_{k_1^{\prime} k_1} \phi_{k_1^{\prime}}(u) \phi_{k_1}(t) = \phi^\top(u) \bm{\rho} \phi(t), \quad \forall u, t \in [0,1], \label{eq:app3} \\
\beta(s,t) &\approx \sum_{k_2=1}^{K_x} \sum_{k_1=1}^{K_y} \beta_{k_2 k_1} \psi_{k_2}(s) \phi_{k_1}(t) = \psi^\top(s) \bm{\beta} \phi(t), \quad \forall s, t \in [0,1], \label{eq:app4}
\end{align} 
where $\phi(t)$ and $\psi(s)$ are the orthonormal SFPC and FPC eigenfunctions, respectively, $y_{i k_1} = \int_0^1 \Y_i(t) \phi_{k_1}(t) dt$, $x_{i k_2} = \int_0^1 \X_i(s) \psi_{k_2}(s) ds$, $\rho_{k_1^{\prime} k_1} = \int_0^1 \int_0^1 \rho(u,t) \phi_{k_1^{\prime}}(u) \phi_{k_1}(t) du dt$, and $\beta_{k_2 k_1} = \int_0^1 \int_0^1 \beta(s,t) \psi_{k_2}(s) \phi_{k_1}(t)$. 

Suppose the error terms $\epsilon_i(t)$, $i \in \{1, \ldots, n\}$, can be decomposed using the SFPC method, utilizing identical sets of orthonormal eigenfunctions within each $\Y_{i}(t)$:
\begin{equation}\label{eq:appe}
\epsilon_{i}(t) = \sum_{k_1=1}^{K_y} e_{i k_1} \phi_{k_1}(t) = \bm{e}_{i}^{\top} \phi(t), \quad \forall t \in [0,1],
\end{equation}
where $e_{i k_1} = \int_0^1 \epsilon_i(t) \phi_{k_1}(t) dt$. Then, substituting~\eqref{eq:app1}-\eqref{eq:appe} in~\eqref{eq:Mform}, the SFoFR model under investigation can be re-expressed as follows:
\begin{equation}\label{eq:redform1}
\bm{Y}^{\top} \phi(t) = \bm{W} \int_0^1 \bm{Y}^{\top} \phi(u) \phi^{\top}(u) \bm{\rho} \phi(t) du + \int_0^1 \bm{X}^{\top} \psi(s) \psi^{\top}(s) \bm{\beta} \phi(t) ds + \bm{e}^{\top} \phi(t),
\end{equation}
where $\bm{Y}^{\top} = [\bm{y}_1^{\top}, \ldots, \bm{y}_n^{\top}]^{\top}$, $\bm{X}^{\top} = [\bm{x}_1^{\top}, \ldots, \bm{x}_n^{\top}]^{\top}$, and $\bm{e}^{\top} = [\bm{e}_1^{\top}, \ldots, \bm{e}_n^{\top}]^{\top}$. By orthonormalities of $\phi(u)$ and $\psi(s)$, i.e., $\int_{0}^{1} \phi(u) \phi^{\top} (u) du = 1$ (similarly for $\psi(s)$, \eqref{eq:redform1} is expressed as:
\begin{equation}\label{eq:redform2}
\bm{Y}^{\top} \phi(t) = \bm{W} \bm{Y}^{\top} \bm{\rho} \phi(t) + \bm{X}^{\top} \bm{\beta} \phi(t) + \bm{e}^{\top} \phi(t).
\end{equation}
Multiplying both sides of~\eqref{eq:redform2} from the right by $\phi^\top(t)$ and integrating with respect to the function support $[0,1]$ yields the reduced form for SFoFR as follows:
\begin{align}
\int_0^1 \bm{Y}^{\top} \phi(t) \phi^{\top}(t) dt &= \bm{W} \int_0^1 \bm{Y}^{\top} \bm{\rho} \phi(t) \phi^{\top}(t) dt + \int_0^1 \bm{X}^{\top} \bm{\beta} \phi(t) \phi^{\top}(t) dt + \int_0^1 \bm{e}^{\top} \phi(t) \phi^{\top}(t) dt, \nonumber \\
\bm{Y}^\top &= \bm{W} \bm{Y}^\top \bm{\rho} + \bm{X}^{\top} \bm{\beta} + \bm{e}^{\top},  \label{eq:redform}
\end{align}
where $\bm{e}_i$, $i \in \{1, \ldots, n\}$ is assumed to have mean $\bm{0}$ and $\mathrm{Cov}(\bm{e}) = \bm{\Sigma}_e \in \mathbb{R}^{K_y \times K_y}$.

The findings suggest that the infinite-dimensional SFoFR model can be represented by projecting it into a finite-dimensional space of FPC and SFPC coefficients. Put differently, estimating the infinite-dimensional SFoFR model becomes equivalent to estimating the multivariate spatial autoregressive (MSAR) model described in~\eqref{eq:redform}. 

\subsection{Least-squares estimation}

To explore the feasible parameter space of~\eqref{eq:redform}, we reformulate the MSAR model into a vectorized representation:
\begin{equation}\label{eq:vecform1}
\widetilde{Y} = (\bm{\rho}^\top \otimes \bm{W}) \widetilde{Y} + \bm{X}^{\top*} \widetilde{\beta} + \widetilde{e},
\end{equation}
where $\widetilde{Y} = \text{vec}(\bm{Y}^\top) = (Y_1, \ldots, Y_{K_y})^\top \in \mathbb{R}^{n K_y}$, $\bm{X}^{\top*} = \mathbb{I}_{K_y} \otimes \bm{X}^{\top}$, $\widetilde{e} = \text{vec}(\bm{e}^\top) = (e_1, \ldots, e_{K_y})^\top \in \mathbb{R}^{n K_y}$, $\widetilde{\beta} = \text{vec}(\bm{\beta}) \in \mathbb{R}^{K_y K_x}$, and $\otimes$ represents the Kronecker product. Let $\lambda_{k_1}(\bm{\rho})$ be the $k_1^\textsuperscript{th}$ eigenvalue of $\bm{\rho}$ such that $\vert \lambda_{1}(\bm{\rho}) \vert \geq \ldots \geq \vert \lambda_{K_y}(\bm{\rho}) \vert$. According to Lemma 1 in \cite{Zhu2020}, the invertibility of the matrix $(\mathbb{I}_{n K_y} - \bm{\rho}^\top \otimes \bm{W})$ is assured under the condition $\vert \lambda_{1}(\bm{\rho}) \vert < 1$. This condition holds, and the vectorized form in~\eqref{eq:vecform1} can therefore be represented as
\begin{align}\label{eq:vecform2}
\widetilde{Y} &= (\mathbb{I}_{n K_y} - \bm{\rho}^\top \otimes \bm{W})^{-1} \left(\bm{X}^{\top*} \widetilde{\beta} + \widetilde{e}\right), \nonumber \\
&= (\mathbb{I}_{n K_y} - \bm{\rho}^\top \otimes \bm{W})^{-1} \left\lbrace \left(\mathbb{I}_{K_y} \otimes \bm{X}^\top \right) \text{vec}(\bm{\beta}) + \widetilde{e} \right\rbrace
\end{align}

For $i, i^{\prime} \in \{1, \ldots, n \}$ and $k_1, k_1^{\prime} \in \{1, \ldots, K_y\}$, let $\bm{Y}_{-(i^{\prime} k_1^{\prime})} = \bm{y}_{ik_1}$, $(i,k_1) \neq (i^{\prime} k_1^{\prime})$, $\bm{Y}^* = \mathbb{E}\{\bm{y}_{ik_1} \vert \bm{Y}_{-(ik_1)}\} \in \mathbb{R}^{n \times K_y}$, $i \in \{1, \ldots, n\}$ and $k_1 \in \{1, \ldots, K_y\}$, and $\widetilde{Y}^* = \text{vec}(\bm{Y}^*)$. In addition, let $\bm{S} = \mathbb{I}_{n K_y} - \bm{\rho}^\top \otimes \bm{W}$ and define $\bm{\Omega} = \bm{\Sigma}^{-1} = \bm{S}^\top(\bm{\Sigma}_{e}^{-1} \otimes \mathbb{I}_n) \bm{S}$, $m = \text{diag}^{-1}(\bm{\Omega}) = \lbrace \text{diag}(\bm{\Sigma}_e^{-1}) \otimes \mathbb{I}_n + \text{diag}(\bm{\rho} \bm{\Sigma}_e^{-1} \bm{\rho}^\top) \otimes \text{diag}(\bm{W}^\top \bm{W}) \rbrace^{-1}$. Then, the following equality exists:
\begin{equation*}
\widetilde{Y}^{*} = \bm{U} - m(\bm{\Omega} - m^{-1}) (\widetilde{Y} - \bm{U}),
\end{equation*}
where $\bm{U} = \mathbb{E}(\widetilde{Y}) = \bm{S}^{-1} \{(\mathbb{I}_{K_y} \otimes \bm{X}^{\top}) \widetilde{\beta}\}$. Let $\widetilde{\rho} = \text{vec}(\bm{\rho}) \in \mathbb{R}^{K_y^2}$ and $\zeta_e = \text{vec}^*(\bm{\Sigma}_e^{-1}) \in \mathbb{R}^{K_y (K_y+1)/2}$, where $\text{vec}^*(\bm{\Sigma}_e^{-1})$ selects the upper triangular part of $\bm{\Sigma}_e^{-1}$. Define $\bm{\Theta} = (\widetilde{\rho}^\top, \widetilde{\beta}^\top, \zeta_e^\top) \in \mathbb{R}^{n*}$ to collect the parameters to be estimated, where $n^* = K_y^2 + K_y K_x + K_y (K_y+1)/2$. Consequently, an objective function of least squares type can be formulated as
\begin{equation}\label{eq:objf}
Q(\bm{\Theta}) = \left\Vert \widetilde{Y} - \widetilde{Y}^{*} \right\Vert^2 = \left\Vert m \bm{S}^\top (\bm{\Sigma}_e^{-1} \otimes \mathbb{I}) (\bm{S} \widetilde{Y} - \bm{X}^{\top*} \widetilde{\beta}) \right\Vert^2.
\end{equation}

\begin{remark}\label{rem:1}
The identification issue of least-squares estimation in the MSAR model~\eqref{eq:vecform1} was discussed by \cite{Zhu2020}. Let $\bm{\rho}_0$, $\widetilde{\beta}_0$, $\bm{\Sigma}_{e_0}$, and $\bm{\Theta}_0$ to be the true parameters. In addition, let $\bm{S}_0 = \mathbb{I}_{n K_y} - \bm{\rho}^\top_0 \otimes \bm{W}$ and $\bm{\Sigma}_0 = \bm{S}_0^{-1} (\bm{\Sigma}_{e_0} \otimes \mathbb{I}_n )(\bm{S}_0^\top)^{-1}$. Then, the expected least squares objective function, denoted by $\mathbb{Q}(\bm{\Theta}) = \mathbb{E} \lbrace Q(\bm{\Theta}) \rbrace$, can be computed as follows:
\begin{equation*}
\mathbb{Q}(\bm{\Theta}) = \Vert m \bm{S}^\top (\bm{\Sigma}_e^{-1} \otimes \mathbb{I})\{\bm{S} \bm{S}_0^{-1} (\mathbb{I}_{K_y} \otimes \bm{X}^{\top}) \widetilde{\beta}_0 - (\mathbb{I}_{K_y} \otimes \bm{X}^{\top}) \widetilde{\beta}\} \Vert^2 + \text{tr}(m \bm{\Omega} \bm{\Sigma}_{0} \bm{\Omega} m).
\end{equation*}
Let $J_{k_1}$, $k_1 \in \{1, \ldots, K_y\}$ denote a $1 \times K_y$ row vector with all zero elements except for the $k_1^\textsuperscript{th}$ entry, which is 1. Define $(\bm{X}_{k_1}^{\dagger})^\top = (J_{k_1} \otimes \mathbb{I}_n) \bm{S}_0^{-1} \lbrace \mathbb{I}_{K_y} \otimes (\bm{W} \bm{X}^\top) \rbrace \widetilde{\beta}_0$, $(\bm{X}^{\dagger})^\top = [(\bm{X}_{1}^{ \dagger})^\top, \ldots, (\bm{X}_{K_y}^{\dagger})^\top] \in \mathbb{R}^{n \times K_y}$, and $(\bm{X}^{\ddagger})^\top = [(\bm{X}^{\dagger})^\top, \bm{X}^{\top}] \in \mathbb{R}^{n \times (K_y + K_x)}$. If we assume that there exists a strictly positive $\iota$ such that $\min_{\vert \lambda_1(\bm{\rho}) \vert \leq 1 - \iota} \lbrace \lambda_{\min}(\bm{S} \bm{S}^\top) \rbrace \geq \tau$, where $\tau$ is a positive constant. Then, according to Theorem 2 in \cite{Zhu2020}, under the condition that $\lim_{n \rightarrow \infty} n^{-1} [\bm{X}^{\ddagger} (\bm{X}^{\ddagger})^\top]$ exists and is non-singular, $\bm{\rho}_0$ and $\widetilde{\beta}_0$ can be uniquely identified in the parameter space.
\end{remark}

Following Remark~\ref{rem:1}, by minimizing~\eqref{eq:objf}, the least squares type estimator $\widehat{\bm{\Theta}} = (\widehat{\widetilde{\rho}}^\top, \widehat{\widetilde{\beta}}^\top, \widehat{\zeta}_e^\top) = \argmin_{\bm{\Theta}} Q(\bm{\Theta})$, where $ \widehat{\widetilde{\beta}} = \text{vec}(\widehat{\bm{\beta}})$, can be obtained. Finally, the estimates of the spatial autocorrelation function and the regression coefficient functions in~\eqref{eq:Mform} can be obtained as follows:
\begin{align}
\widehat{\rho}(u,t) &= \phi^\top(u) \widehat{\bm{\rho}} \phi(t), \label{eq:estrho} \\
\widehat{\beta}(s,t) &= \psi^\top(s) \widehat{\bm{\beta}} \phi(t). \label{eq:estbeta}
\end{align}
The following theorem discusses the consistency and asymptotic normality properties of the proposed estimators $\rho(u,t)$ and $\beta(s,t)$.
\begin{theorem}\label{th:1}
Let $\bm{\theta} = \{\rho(u,t), \beta(s,t) \}$ denote the true parameters of the SFoFR model and let $\widehat{\bm{\theta}} = \{ \widehat{\rho}(u,t), \widehat{\beta}(s,t) \}$ denote the least squares estimates of $\bm{\theta}$ as in~\eqref{eq:estrho} and~\eqref{eq:estbeta}. Assume the conditions $C_1$-$C_9$ given in Appendix hold. Then, $\widehat{\bm{\theta}}$ is $\sqrt{n}$-consistent and we have:
\begin{equation*}
\sqrt{n} \big( \widehat{\bm{\theta}} - \bm{\theta} \big) \xrightarrow{d} \mathcal{GP} \big( \bm{0}, \bm{\Sigma}_{\bm{\theta}} \big),
\end{equation*}
where $\bm{\Sigma}_{\bm{\theta}}$ is the covariance operator of the asymptotic Gaussian process ($\mathcal{GP}$), represented in terms of the SFPC and FPC eigenfunctions and the finite-dimensional covariance matrices. The detailed representation of $\bm{\Sigma}_{\bm{\theta}}$ is given in the Appendix.
\end{theorem}
The proof of Theorem~\ref{th:1} is postponed to the online supplementary material.

\section{Simulation study}\label{sec:4}

We conduct a series of Monte-Carlo experiments to evaluate the estimation accuracy and predictive performance of the proposed SFoFR method. The finite-sample behavior of our approach is compared against two alternative FoFR methods: one based on the FPC decomposition and another utilizing the functional partial least squares (FPLS) technique as developed by \cite{Zhou2021}.

In the Monte-Carlo experiments, both the functional predictor $\X_i(s)$ and functional response $\Y_i(t)$ are generated at 101 equally spaced points over the interval  $[0,1]$, with $s, t = r/101 \in \{r = 1, \ldots, 101\}$. The functional predictor $\X_i(s)$ is constructed as follows:
\begin{equation*}
\X_i(s) = \sum_{k=1}^{10} \frac{1}{k^{3/2}} \left\{ \nu _{i1,k} \sqrt{2} \cos \left ( k \pi s \right ) + \nu _{i2,k} \sqrt{2} \sin \left ( k \pi s \right ) \right\},
\end{equation*}
where $\nu_{i1,k}$ and $\nu_{i2,k}$ are independent standard normal random variables. The bivariate regression coefficient function $\beta(s,t)$ is defined as:
\begin{equation*}
\beta(s,t) = 2 + s + t + 0.5 \sin(2 \pi s t).
\end{equation*}
The spatial autocorrelation parameter function $\rho(u,t)$ is specified by:
\begin{equation*}
\rho(u,t) = \alpha \frac{1 + u t}{1 + \vert u - t \vert},
\end{equation*}
where $\alpha \in (0, 1)$ regulates the strength of spatial dependence in the data. When $\alpha$ is close to zero, the spatial effect is negligible, indicating minimal correlation among observations. As $\alpha$ approaches 1, the spatial dependence is maximized, signifying a strong influence from neighboring observations. To evaluate different levels of spatial dependence, three values of $\alpha$ are considered: $\alpha \in \{0.1, 0.5, 0.9\}$, corresponding to scenarios with weak, moderate, and strong spatial effects, respectively. The spatial correlation is weak at $\alpha = 0.1$, indicating limited interaction among observations. For $\alpha = 0.5$, the data exhibits a moderate spatial effect, while at $\alpha = 0.9$, the spatial dependence is pronounced, reflecting a high level of influence from adjacent observations.

In our experiments, we consider two approaches for generating the spatial weight matrix $\bm{W}$:
\begin{asparaenum}
\item[1)] The inverse distance weight matrix is constructed such that each entry $w_{i i^{\prime}}$ represents the inverse of the distance between locations $i$ and $i^{\prime}$. Specifically, if $i \neq i^{\prime}$, the weight is defined as:
\begin{equation*}
w_{i i^{\prime}} = \frac{1}{1 + \vert i - i^{\prime} \vert}.
\end{equation*}
The diagonal elements, $w_{i i}$, are set to zero for all $i$. To ensure that each row of $\bm{W}$ sums to one, the matrix is row-normalized:
\begin{equation*}
\bm{W}_{i.} = \frac{w_{i i^{\prime}}}{\sum_{i^{\prime}=1}^n w_{i i^{\prime}}}, \quad \forall i.
\end{equation*}
\item[2)] The exponential distance weight matrix, where each entry $w_{i i^{\prime}}$ decreases exponentially with the distance between locations $i$ and $i^{\prime}$. For $i \neq i^{\prime}$, the weight is defined as:
\begin{equation*}
w_{i i^{\prime}} = \exp(-d \vert i - i^{\prime} \vert),
\end{equation*}
where $d > 0$ is a decay parameter that governs the rate of exponential decline. The diagonal elements $w_{i i}$ are also set to zero, and the matrix is row-normalized, akin to the inverse distance matrix. In our experiments, we set $d = 0.5$. 
\end{asparaenum}
These two matrices represent different assumptions about the spatial dependence structure. The inverse distance weight matrix imposes a linear decay of influence, where the spatial weights decrease gradually as the distance between locations increases, capturing scenarios where the spatial effect diminishes slowly, such as in social or economic interactions that persist over larger distances. In contrast, the exponential distance weight matrix models a more rapid decay in spatial influence, which is suitable for contexts where the effect diminishes sharply with distance, such as in environmental studies where factors like pollutants or temperature effects decrease quickly away from the source. Thus, choosing between these matrices allows for flexibility in modeling varying spatial dynamics depending on the application context.

\begin{figure}[!htb]
\centering
\includegraphics[width=8.5cm]{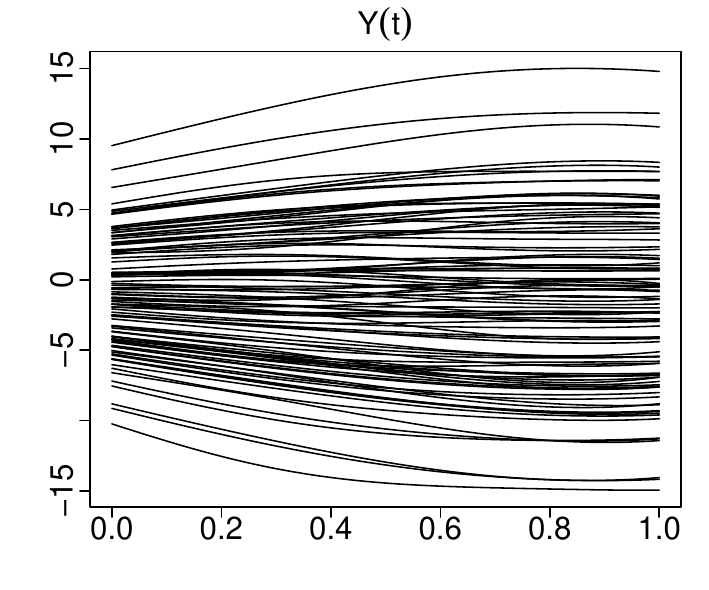}
\qquad
\includegraphics[width=8.5cm]{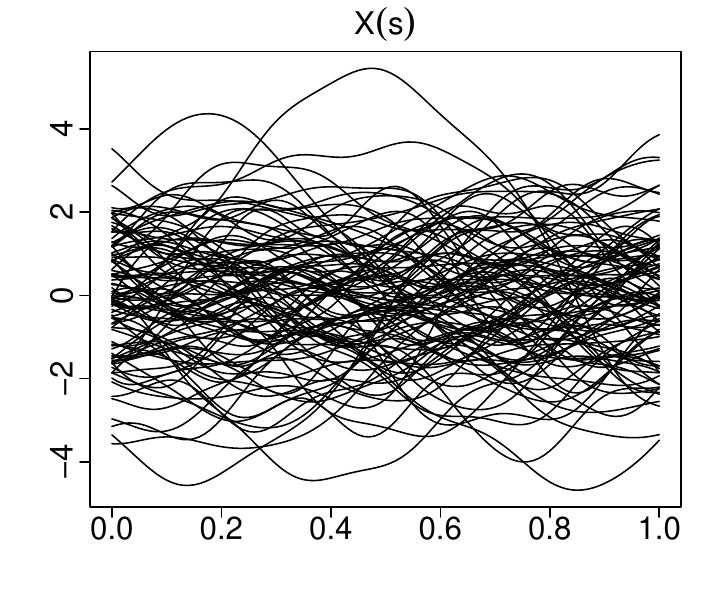}
\\  
\includegraphics[width=8.5cm]{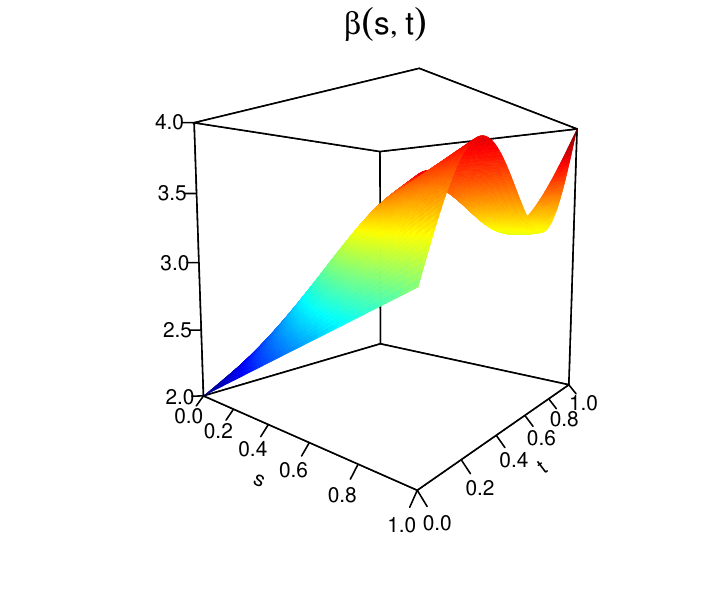}
\qquad
\includegraphics[width=8.5cm]{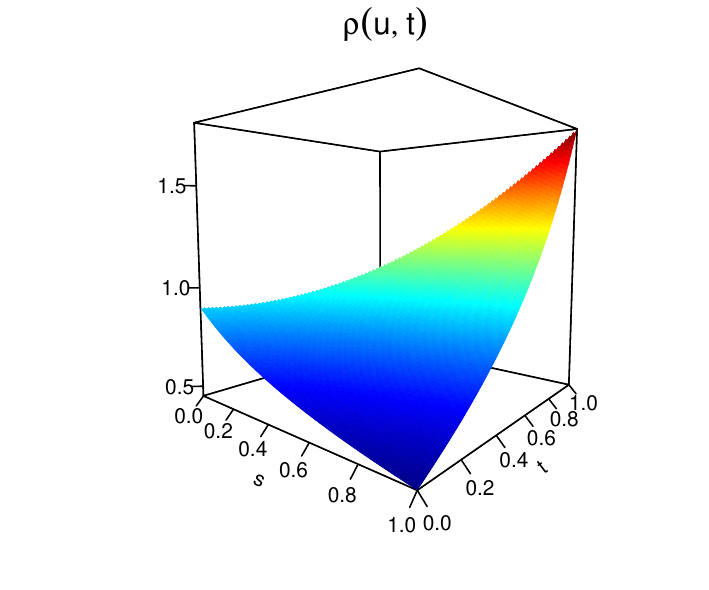}
\caption{\small{Graphs depicting 100 generated sample curves for functional response (top left panel) and functional covariate (top right panel), along with true bivariate regression coefficient functions (bottom left panel) and true spatial autocorrelation parameter function (bottom right panel). The data are generated with exponential distance weight matrix and when $\alpha = 0.9$}.}\label{fig:Fig_1}
\end{figure}

To generate the functional response, we utilize the Neumann series approximation as described in \cite{Hoshino2024}, taking advantage of the fact that the spatial autocorrelation function $\rho(u,t)$ satisfies the contraction condition specified in Proposition~\ref{prop1}. Let $\Y_i(t)$ denote the functional response at location $i$. The response is defined by $\Y_i(t) = G_i(t) + \sum_{k=1}^{\infty} \bm{W}^k G_i(t)$, where $G_i(t) = \int_0^1 \X_i(s) \beta(s,t) ds + \epsilon_i(t)$, and $\epsilon_i(t)$ is a noise term generated from a standard normal distribution. The Neumann series approximation is employed to compute this series until convergence. Since $\rho(u,t)$ satisfies $\Vert \rho \Vert_{\infty} < \frac{1}{\Vert \bm{W} \Vert_{\infty}}$, as required by Proposition~\ref{prop1}, the operator $(\mathbb{I}_d - \mathcal{T})$ is a contraction mapping, ensuring that the Neumann series converges to a unique solution $\Y_i(t) = (\mathbb{I}_d - \mathcal{T})^{-1} G_i(t)$. To accurately capture the spatial effects in the functional response, we iteratively apply the operator $(\mathbb{I}_d - \mathcal{T})$ to the initial response $G_i(t)$. The convergence of the Neumann series is assessed by the approximation $\Y_i(t) \approx \Y_i^{(\mathcal{M})}(t):= \sum_{m = 0}^{\mathcal{M}} \mathcal{T}^{m} G_i(t)$ where $\mathcal{M}$ is incremented until the condition $\max_{i \in \{1, \ldots, n\}} \vert \Y_i^{(\mathcal{M})}(t) - \Y_i^{(\mathcal{M}-1)}(t) \vert < 0.001$ is satisfied for all $t$. This approach ensures the generated functional response accurately reflects the intended spatial correlation structure. Figure~\ref{fig:Fig_1} presents a graphical display of the generated random curves and parameter functions.

In the simulation experiments, we consider four different training sample sizes: $n_{\text{train}} \in \{100, 250$, $500, 1000\}$. For each training dataset, models are constructed. The estimation performance of the proposed method is assessed by calculating the integrated squared errors (ISE) for the bivariate regression coefficient function $\beta(s,t)$ and the spatial autocorrelation parameter function $\rho(u,t)$:
\begin{align*}
\text{ISE}(\widehat{\beta}) &= \int_0^1 \int_0^1 \{\widehat{\beta}(s,t) - \beta(s,t)\}^2 ds dt, \\ 
\text{ISE}(\widehat{\rho}) &= \int_0^1 \int_0^1 \{\widehat{\rho}(u,t) - \rho(u,t)\}^2 du dt,
\end{align*}
where $\widehat{\beta}(s,t)$ and $\widehat{\rho}(u,t)$ are the estimated regression and spatial autocorrelation functions, respectively. For the FPC and FPLS methods, only $\text{ISE}(\widehat{\beta})$ is computed. 

To rigorously assess the predictive performance of the proposed models, we generate an independent testing dataset comprising $n_{\text{test}} = 1000$ samples for each training set size. The fitted models are then evaluated on these test samples. Specifically, we compute the following mean squared error (MSE) and mean squared prediction error (MSPE) metrics to evaluate the in-sample and out-of-sample predictive performance of the methods:
\begin{align*}
\text{MSE} &= \int_0^1 \{\widehat{\Y}(t) - Y(t)\}^2 dt, \\ \text{MSPE} &= \int_0^1 \{\widehat{\Y}^*(t) - Y^*(t)\}^2 dt,
\end{align*}
where $Y(t)$ and $Y^*(t)$ denote the response variables for the training and test datasets, respectively. Correspondingly, $\widehat{Y}(t)$ and $\widehat{\Y}^*(t)$ represent the model predictions for the training and test samples. In all cases, the training and test samples are mutually exclusive to ensure that out-of-sample predictions are obtained without bias from shared data points. If there are common points between training and test sets, a trend-corrected strategy, such as that proposed by \cite{Goulard2017}, may be applied to achieve unbiased out-of-sample predictions.

For the proposed method and the FPC-based approach, the number of principal components is selected to capture at least 95\% of the total variance in the data. In contrast, the number of components for the FPLS method is determined using a cross-validation procedure, as recommended by \cite{Zhou2021}. This strategy ensures that each method is tuned optimally for accurate estimation and prediction performance. For all training sample sizes and spatial weight matrix types, we perform 500 independent Monte-Carlo experiments.

The computed mean $\text{ISE}(\widehat{\beta})$, $\text{ISE}(\widehat{\rho})$, MSE, and MSPE values with their standard errors are presented in Table~\ref{tab:tab_1} and~\ref{tab:tab_2}. The results in Table~\ref{tab:tab_1} correspond to data generated using an inverse distance weight matrix, while Table~\ref{tab:tab_2} presents the results for data generated using an exponential distance weight matrix. When the data are generated using inverse distance weight matrix, that is Table~\ref{tab:tab_1}, for weak spatial dependence ($\alpha = 1$), the FPLS method yields the lowest ISE for the estimated regression coefficient function outperforming the SFoFR and FPC methods across all sample sizes. This suggests that in cases of minimal spatial autocorrelation, the FPLS efficiently captures the underlying data structure. However, as spatial dependence increases to moderate levels ($\alpha = 0.5$), the SFoFR model shows its strengths. In this case, the FPLS still produces improved ISE values over the FPC and proposed SFoFR model when the training sample size is small (i.e., $n_{\text{train}} \in\{ 100, 250\}$, while the proposed method produces improved $\text{ISE}(\widehat{\beta})$ values over the FPLS and FPC methods. In the strong spatial dependence scenario ($\alpha = 0.9)$, the SFoFR model exhibits a clear advantage. It achieves the smallest ISE for $\beta(s,t)$, particularly at larger sample sizes, where the SFoFR model demonstrates robustness in parameter estimation. The MSE and MSPE for SFoFR are also substantially lower than that of FPC and FPLS, highlighting its superior predictive performance when spatial dependencies are strong. These results indicate that the inverse distance weight matrix amplifies the spatial relationships captured by SFoFR, allowing it to outperform competing methods in this context.
\begin{small}
\begin{center}
\tabcolsep 0.315in
\begin{longtable}{@{}cccccccc@{}} 
\caption{\small{Computed mean $\text{ISE}(\widehat{\beta})$, $\text{ISE}(\widehat{\rho})$, MSE, and MSPE values with their standard errors (given in brackets) over 500 Monte-Carlo replications when the data are generated using inverse distance weight matrix. The results are obtained under four sample sizes ($n_{\text{train}}$)}.}\label{tab:tab_1} \\
\toprule
{$\alpha$} & {$n_{\text{train}}$} & Method & $\text{ISE}(\widehat{\beta})$ & $\text{ISE}(\widehat{\rho})$ & MSE & MSPE \\ \midrule
\endfirsthead
\toprule
{$\alpha$} & {$n_{\text{train}}$} & Method & $\text{ISE}(\widehat{\beta})$ & $\text{ISE}(\widehat{\rho})$ & MSE & MSPE \\ \midrule
\endhead
\midrule
\multicolumn{7}{r}{Continued on next page} \\ 
\endfoot
\endlastfoot
0.1 & 100   & FPLS  & $0.027$   & $--$   & $0.001$    & $0.091$\\
    &       &       & ($0.010$) & ($--$) & ($0.001$) & ($0.099$) \\
    &       & FPC   & $0.147$   & $--$   & $0.051$    & $0.054$  \\
    &       &       & ($0.124$) & ($--$) & ($0.010$)  & ($0.010$) \\
    &       & SFoFR & $0.233$   & $0.027$  & $0.097$    & $0.115$  \\
    &       &       & ($0.218$) & ($0.091$) & ($0.084$) & ($0.108$) \\
\cmidrule(l){2-7}				
    & 250   & FPLS  & $0.016$   & $--$      & $0.001$   & $0.029$\\
    &       &       & ($0.005$) & ($--$)    & ($<0.001$) & ($0.032$)  \\
    &       & FPC   & $0.091$   & $--$      & $0.048$ & $0.048$  \\
    &       &       & ($0.046$) & ($--$) & ($0.005$) & ($0.004$) \\
    &       & SFoFR & $0.138$ & $0.041$ & $0.077$ & $0.081$  \\
    &       &       & ($0.090$) & ($0.176$) & ($0.058$) & ($0.062$) \\
\cmidrule(l){2-7}
& 500   & FPLS  & $0.013$   & $--$   & $0.001$    & $0.013$\\
&       &       & ($0.002$) & ($--$) & ($<0.001$) & ($0.013$)  \\

&       & FPC   & $0.079$   & $--$   & $0.046$   & $0.047$  \\
&       &       & ($0.028$) & ($--$) & ($0.004$) & ($0.003$) \\

&       & SFoFR & $0.139$   & $0.064$   & $0.092$   & $0.093$  \\
&       &       & ($0.090$) & ($0.120$) & ($0.060$) & ($0.060$) \\
\cmidrule(l){2-7}

& 1000  & FPLS  & $0.012$   & $--$   & $0.001$   & $0.006$\\
&       &       & ($0.002$) & ($--$) & ($<0.001$) & ($0.005$)  \\

&       & FPC   & $0.064$ & $--$ & $0.046$ & $0.046$  \\
&       &       & ($0.026$) & ($--$) & ($0.001$) & ($0.001$) \\

&       & SFoFR & $0.120$ & $0.007$ & $0.092$ & $0.094$  \\
&       &       & ($0.084$) & ($0.025$) & ($0.071$) & ($0.074$) \\
\midrule

0.5 & 100   & FPLS  & $0.204$   & $--$  & $0.056$ & $0.319$\\
    &       &       & ($0.125$) & ($--$) & ($0.025$) & ($0.308$)  \\

    &       & FPC   & $0.141$   & $--$   & $0.109$ & $0.162$  \\
    &       &       & ($0.102$) & ($--$) & ($0.030$) & ($0.082$) \\

    &       & SFoFR & $0.256$ & $0.081$ & $0.158$ & $0.171$  \\
    &       &       & ($0.386$) & ($0.304$) & ($0.291$) & ($0.203$) \\
\cmidrule(l){2-7}
					
& 250   & FPLS  & $0.092$   & $--$      & $0.048$   & $0.146$\\
&       &       & ($0.055$) & ($--$)    & ($0.016$) & ($0.121$)  \\

&       & FPC   & $0.090$ & $--$ & $0.095$ & $0.114$  \\
&       &       & ($0.039$) & ($--$) & ($0.017$) & ($0.038$) \\

&       & SFoFR & $0.145$ & $0.028$ & $0.090$ & $0.097$  \\
&       &       & ($0.084$) & ($0.099$) & ($0.064$) & ($0.070$) \\
\cmidrule(l){2-7}

& 500   & FPLS  & $0.063$   & $--$   & $0.039$   & $0.093$\\
&       &       & ($0.036$) & ($--$) & ($0.011$) & ($0.064$)  \\

&       & FPC   & $0.076$   & $--$   & $0.085$   & $0.099$  \\
&       &       & ($0.024$) & ($--$) & ($0.012$) & ($0.024$) \\

&       & SFoFR & $0.130$  & $0.070$   & $0.085$   & $0.089$  \\
&       &       & ($0.068$) & ($0.013$) & ($0.044$) & ($0.048$) \\
\cmidrule(l){2-7}

& 1000 & FPLS   & $0.042$ & $--$ & $0.032$ & $0.064$\\
&       &       & ($0.018$) & ($--$) & ($0.006$) & ($0.034$)  \\

&       & FPC   & $0.066$ & $--$ & $0.076$ & $0.092$  \\
&       &       & ($0.015$) & ($--$) & ($0.007$) & ($0.017$) \\

&       & SFoFR & $0.039$ & $0.010$ & $0.074$ & $0.075$  \\
&       &       & ($0.042$) & ($<0.001$) & ($0.025$) & ($0.025$) \\
\midrule

0.9     & 100   & FPLS & $0.596$ & $--$ & $0.488$ & $8.122$\\
&       &       & ($0.717$) & ($--$) & ($0.251$) & ($8.598$)  \\

&       & FPC   & $0.227$ & $--$ & $0.466$ & $6.813$  \\
&       &       & ($0.140$) & ($--$) & ($0.253$) & ($7.118$) \\

&       & SFoFR & $0.240$ & $0.035$ & $0.179$ & $0.524$  \\
&       &       & ($0.158$) & ($0.008$) & ($0.118$) & ($0.759$) \\
\cmidrule(l){2-7}
					
& 250   & FPLS & $0.298$ & $--$ & $0.313$ & $4.021$\\
&       &       & ($0.210$) & ($--$) & ($0.146$) & ($4.392$)  \\

&       & FPC   & $0.189$ & $--$ & $0.367$ & $3.601$  \\
&       &       & ($0.053$) & ($--$) & ($0.150$) & ($3.803$) \\

&       & SFoFR & $0.130$ & $0.032$ & $0.096$ & $0.122$  \\
&       &       & ($0.060$) & ($0.001$) & ($0.031$) & ($0.093$) \\
\cmidrule(l){2-7}

& 500   & FPLS  & $0.180$ & $--$ & $0.274$ & $2.463$\\
&       &       & ($0.111$) & ($--$) & ($0.116$) & ($2.531$)  \\

&       & FPC   & $0.125$ & $--$ & $0.327$ & $1.749$  \\
&       &       & ($0.040$) & ($--$) & ($0.117$) & ($1.564$) \\

&       & SFoFR & $0.099$ & $0.030$ & $0.080$ & $0.088$  \\
&       &       & ($0.033$) & ($0.001$) & ($0.016$) & ($0.024$) \\
\cmidrule(l){2-7}

& 1000  & FPLS  & $0.093$ & $--$ & $0.216$ & $1.564$\\
&       &       & ($0.059$) & ($--$) & ($0.075$) & ($1.385$)  \\

&       & FPC   & $0.091$ & $--$ & $0.296$ & $1.251$  \\
&       &       & ($0.027$) & ($--$) & ($0.075$) & ($1.090$) \\

&       & SFoFR & $0.085$ & $0.028$ & $0.071$ & $0.074$  \\
&       &       & ($0.019$) & ($<0.001$) & ($0.010$) & ($0.012$) \\
\bottomrule
\end{longtable}
\end{center}
\end{small}

\begin{small}
\begin{center}
\tabcolsep 0.315in
\renewcommand{\arraystretch}{0.9}
\begin{longtable}{@{}cccccccc@{}} 
\caption{\small{Computed mean $\text{ISE}(\widehat{\beta})$, $\text{ISE}(\widehat{\rho})$, MSE, and MSPE values with their standard errors (given in brackets) over 500 Monte-Carlo replications when the data are generated using exponential distance weight matrix. The results are obtained under four sample sizes ($n_{\text{train}}$)}.}\label{tab:tab_2} \\
\toprule
{$\alpha$} & {$n_{\text{train}}$} & Method & $\text{ISE}(\widehat{\beta})$ & $\text{ISE}(\widehat{\rho})$ & MSE & MSPE \\ \midrule
\endfirsthead
\toprule
{$\alpha$} & {$n_{\text{train}}$} & Method & $\text{ISE}(\widehat{\beta})$ & $\text{ISE}(\widehat{\rho})$ & MSE & MSPE \\ \midrule
\endhead
\midrule
\multicolumn{7}{r}{Continued on next page} \\ 
\endfoot
\endlastfoot
0.1 & 100   & FPLS  & $0.070$   & $--$  & $0.010$ & $0.081$\\
    &       &       & ($0.037$) & ($--$) & ($0.002$) & ($0.074$)  \\

    &       & FPC   & $0.154$ & $--$ & $0.062$ & $0.067$  \\
    &       &       & ($0.118$) & ($--$) & ($0.010$) & ($0.009$) \\

    &       & SFoFR & $0.253$ & $0.031$ & $0.100$ & $0.106$  \\
    &       &       & ($0.128$) & ($0.132$) & ($0.079$) & ($0.081$) \\
\cmidrule(l){2-7}
					
& 250   & FPLS  & $0.050$   & $--$   & $0.011$   & $0.046$\\
&       &       & ($0.019$) & ($--$) & ($0.001$) & ($0.035$)  \\

&       & FPC   & $0.097$ & $--$ & $0.058$ & $0.064$  \\
&       &       & ($0.055$) & ($--$) & ($0.004$) & ($0.003$) \\

&       & SFoFR & $0.158$ & $0.040$ & $0.093$ & $0.095$  \\
&       &       & ($0.135$) & ($0.176$) & ($0.065$) & ($0.067$) \\
\cmidrule(l){2-7}

& 500   & FPLS  & $0.033$ & $--$ & $0.011$ & $0.025$\\
&       &       & ($0.013$) & ($--$) & ($0.001$) & ($0.014$)  \\

&       & FPC   & $0.070$ & $--$ & $0.056$ & $0.058$  \\
&       &       & ($0.020$) & ($--$) & ($0.003$) & ($0.002$) \\

&       & SFoFR & $0.113$ & $0.009$ & $0.088$ & $0.088$  \\
&       &       & ($0.071$) & ($0.031$) & ($0.058$) & ($0.060$) \\
\cmidrule(l){2-7}

& 1000  & FPLS  & $0.025$   & $--$ & $0.012$ & $0.017$\\
&       &       & ($0.008$) & ($--$) & ($<0.001$) & ($0.005$)  \\

&       & FPC   & $0.066$ & $--$ & $0.056$ & $0.057$  \\
&       &       & ($0.024$) & ($--$) & ($0.002$) & ($0.002$) \\

&       & SFoFR & $0.099$ & $<0.001$ & $0.079$ & $0.079$  \\
&       &       & ($0.053$) & ($<0.001$) & ($0.040$) & ($0.041$) \\
\midrule

0.5 & 100 & FPLS & $0.897$ & $--$ & $0.586$ & $0.977$\\
    & & & ($1.572$) & ($--$) & ($0.192$) & ($0.262$)  \\

& & FPC & $0.246$ & $--$ & $0.651$ & $0.804$  \\
& & & ($0.142$) & ($--$) & ($0.178$) & ($0.106$) \\

& & SFoFR & $0.182$ & $0.010$ & $0.072$ & $0.077$  \\
& & & ($0.111$) & ($<0.001$) & ($0.020$) & ($0.021$) \\
\cmidrule(l){2-7}
					
& \multirow{6}{*}{250} & FPLS & $0.448$ & $--$ & $0.607$ & $0.770$\\
& & & ($0.265$) & ($--$) & ($0.127$) & ($0.139$)  \\

& & FPC & $0.146$ & $--$ & $0.663$ & $0.739$  \\
& & & ($0.071$) & ($--$) & ($0.127$) & ($0.077$) \\

& & SFoFR & $0.116$ & $0.009$ & $0.065$ & $0.066$  \\
& & & ($0.039$) & ($<0.001$) & ($0.010$) & ($0.009$) \\
\cmidrule(l){2-7}

& \multirow{6}{*}{500} & FPLS & $0.369$ & $--$ & $0.627$ & $0.715$\\
& & & ($0.158$) & ($--$) & ($0.091$) & ($0.096$)  \\

& & FPC & $0.122$ & $--$ & $0.682$ & $0.722$  \\
& & & ($0.050$) & ($--$) & ($0.091$) & ($0.069$) \\

& & SFoFR & $0.095$ & $0.010$ & $0.063$ & $0.064$  \\
& & & ($0.026$) & ($<0.001$) & ($0.007$) & ($0.006$) \\
\cmidrule(l){2-7}

& \multirow{6}{*}{1000} & FPLS & $0.251$ & $--$ & $0.636$ & $0.685$\\
& & & ($0.185$) & ($--$) & ($0.068$) & ($0.078$)  \\

& & FPC & $0.103$ & $--$ & $0.690$ & $0.713$  \\
& & & ($0.032$) & ($--$) & ($0.069$) & ($0.068$) \\

& & SFoFR & $0.090$ & $0.009$ & $0.062$ & $0.062$  \\
& & & ($0.015$) & ($<0.001$) & ($0.004$) & ($0.004$) \\
\midrule

0.9 & 100 & FPLS & $7.066$ & $--$ & $17.400$ & $32.374$\\
& & & ($9.685$) & ($--$) & ($8.391$) & ($9.145$)  \\

& & FPC & $3.473$ & $--$ & $17.486$ & $30.390$  \\
& & & ($2.590$) & ($--$) & ($8.375$) & ($7.510$) \\

& & SFoFR & $0.201$ & $0.030$ & $0.166$ & $0.207$  \\
& & & ($0.113$) & ($<0.001$) & ($0.072$) & ($0.121$) \\
\cmidrule(l){2-7}
					
& 250 & FPLS & $5.164$ & $--$ & $20.554$ & $27.467$\\
& & & ($5.568$) & ($--$) & ($5.998$) & ($5.337$)  \\

& & FPC & $2.860$ & $--$ & $20.687$ & $26.631$  \\
& & & ($2.117$) & ($--$) & ($6.025$) & ($4.850$) \\

& & SFoFR & $0.143$ & $0.030$ & $0.102  $ & $0.107$  \\
& & & ($0.038$) & ($<0.001$) & ($0.019$) & ($0.023$) \\
\cmidrule(l){2-7}

& 500 & FPLS & $3.639$ & $--$ & $22.129$ & $26.058$\\
& & & ($2.364$) & ($--$) & ($5.226$) & ($4.610$)  \\

& & FPC & $2.374$ & $--$ & $22.235$ & $25.642$  \\
& & & ($1.206$) & ($--$) & ($5.223$) & ($4.364$) \\

& & SFoFR & $0.128$ & $0.030$ & $0.092$ & $0.094$  \\
& & & ($0.021$) & ($<0.001$) & ($0.011$) & ($0.010$) \\
\cmidrule(l){2-7}

& 1000 & FPLS & $3.275$ & $--$ & $22.717$ & $24.462$\\
& & & ($2.798$) & ($--$) & ($3.807$) & ($4.060$)  \\

& & FPC & $2.278$ & $--$ & $22.820$ & $24.295$  \\
& & & ($0.840$) & ($--$) & ($3.813$) & ($3.934$) \\

& & SFoFR & $0.120$ & $0.030$ & $0.090$ & $0.090$  \\
& & & ($0.014$) & ($<0.001$) & ($0.007$) & ($0.007$) \\
\bottomrule
\end{longtable}
\end{center}
\end{small}

When the data are generated using exponential distance weight matrix, that is Table~\ref{tab:tab_2}, when the spatial dependence is weak ($\alpha = 1$), the FPLS method again provides the best performance in terms of ISE for $\beta(s,t)$, particularly for smaller sample sizes. This is likely because the exponential decay of the spatial weights limits the impact of distant neighbors, making the problem less reliant on the spatial structure and favoring FPLS, which does not directly account for spatial autocorrelation. As spatial dependence increases $(\alpha = 0.5)$, the SFoFR model performs competitive ISE for $\beta(s,t)$, especially for larger sample sizes. This suggests that SFoFR effectively captures the exponential decay structure as the model leverages the localized nature of spatial relationships. For strong spatial dependence ($\alpha = 0.9)$, the SFoFR model once again outperforms both FPLS and FPC across all metrics. MSPE for SFoFR is significantly lower than for the other methods, particularly for large training sample sizes. These findings suggest the SFoFR model is well-suited for scenarios with strong localized spatial effects, further highlighting its adaptability to different spatial weight matrices.

For both cases, the proposed SFoFR model produces smaller ISE values for $\rho(u,t)$, highlighting the model’s robustness in capturing strong, localized spatial dependencies. Thus, the proposed method produces significantly improved MSE and MSPE results over the FPLS and FPC methods, particularly when the data present a strong spatial correlation. 

A comparison of the results from Tables~\ref{tab:tab_1} and~\ref{tab:tab_2} reveals that the choice of spatial weight matrix influences the performance of the SFoFR model. When the inverse distance weight matrix is used, the SFoFR model excels in scenarios with strong spatial dependence, capturing long-range spatial relationships effectively. On the other hand, the exponential distance weight matrix tends to favor more localized spatial effects, and the SFoFR model performs exceptionally well in such settings. For the estimation of bivariate regression coefficient function $\beta(s,t)$, the FPLS method performs consistently better than SFoFR and FPC when spatial dependencies are weak, as it does not rely on spatial structure for dimension reduction. 

The SFoFR model demonstrates considerable flexibility and adaptability across different spatial weight matrices. It offers superior estimation and predictive performance in scenarios where spatial dependencies are moderate to strong, regardless of whether those dependencies are long-range inverse distance weight matrix or localized exponential distance weight matrix. These results underscore the robustness of the SFoFR model in handling various spatial structures, making it a powerful tool for spatially dependent functional data.

\section{Application to Brazilian COVID-19 data}\label{sec:5}

We consider a dataset comprising COVID-19 statistics from 5,570 Brazilian cities, allowing for an in-depth examination of spatial dependencies in confirmed cases and deaths across the country. The data, which was sourced utilizing the \texttt{COVID19} \Rlogo\ package \citep{COVID19}, encapsulates both the average confirmed cases and average deaths for the years 2021 and 2022, facilitating a nuanced understanding of the intercity relationships.

Recently, a variety of regression models have been developed to analyze COVID-19 datasets, shedding light on the relationship between fatalities resulting from the virus and a range of influencing factors \citep[see, e.g.,][]{Giordano21, Acal21, BSM2024}. Despite these advancements, existing models often overlook the critical aspect of spatial dependence among the response variables. This omission limits our understanding of how the geographic proximity of cities may influence COVID-19 prevalence and outcomes. Addressing this gap is essential for capturing the complex interactions that characterize the spread of the virus and its impact across different regions. 

\begin{figure}[!htb]
\centering
\includegraphics[width=8.6cm]{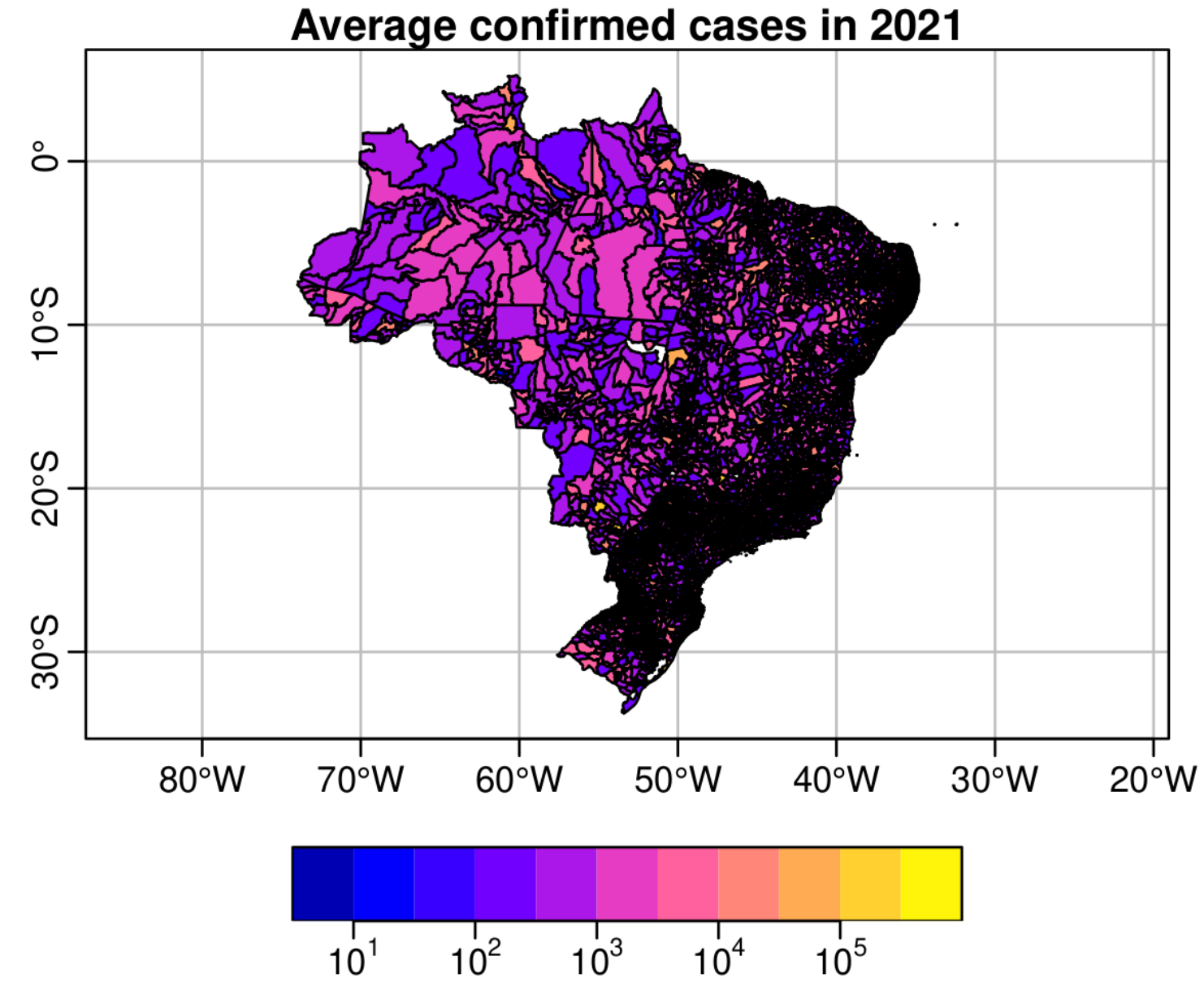}
\qquad
\includegraphics[width=8.6cm]{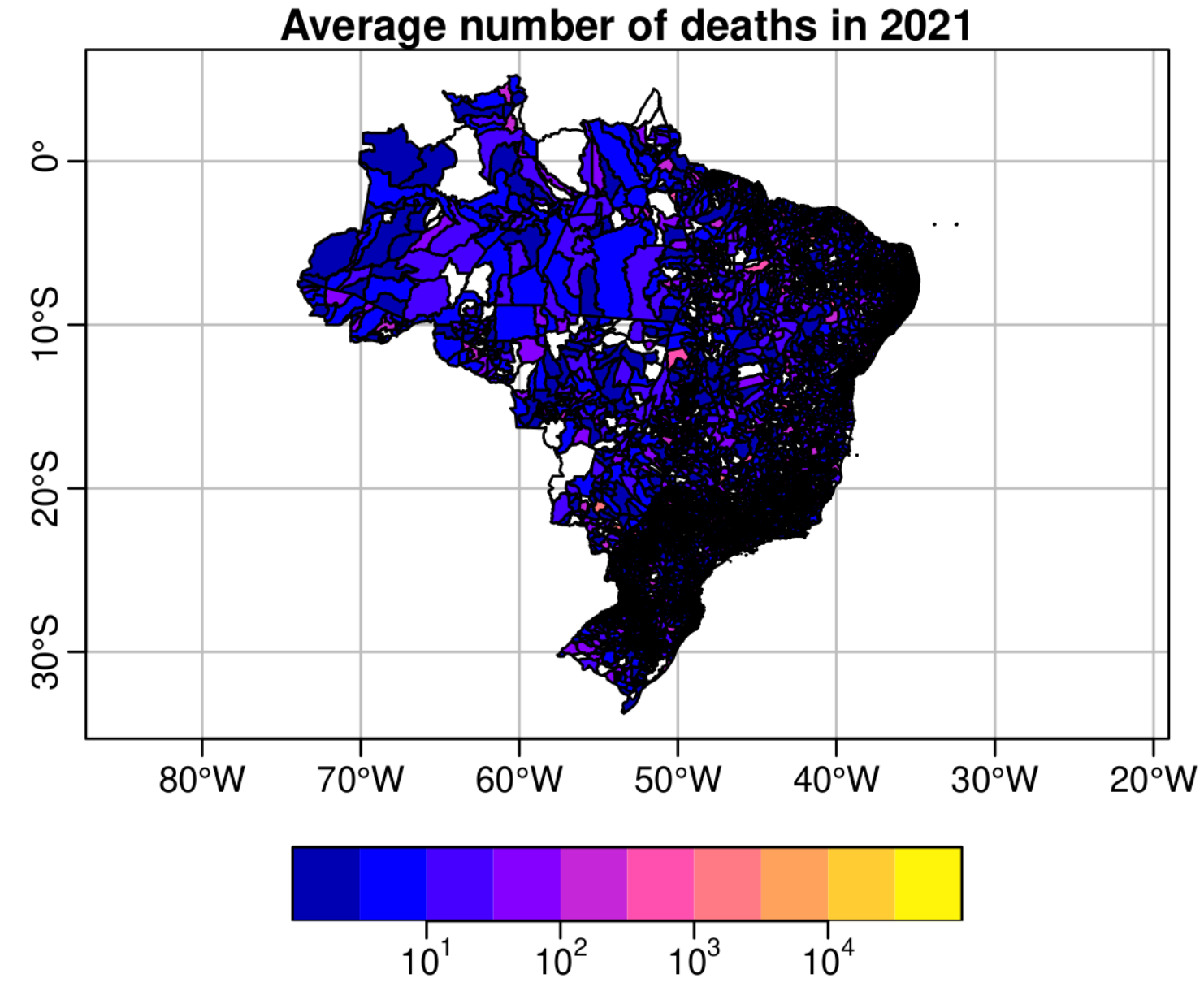}
\\ 
\includegraphics[width=8.6cm]{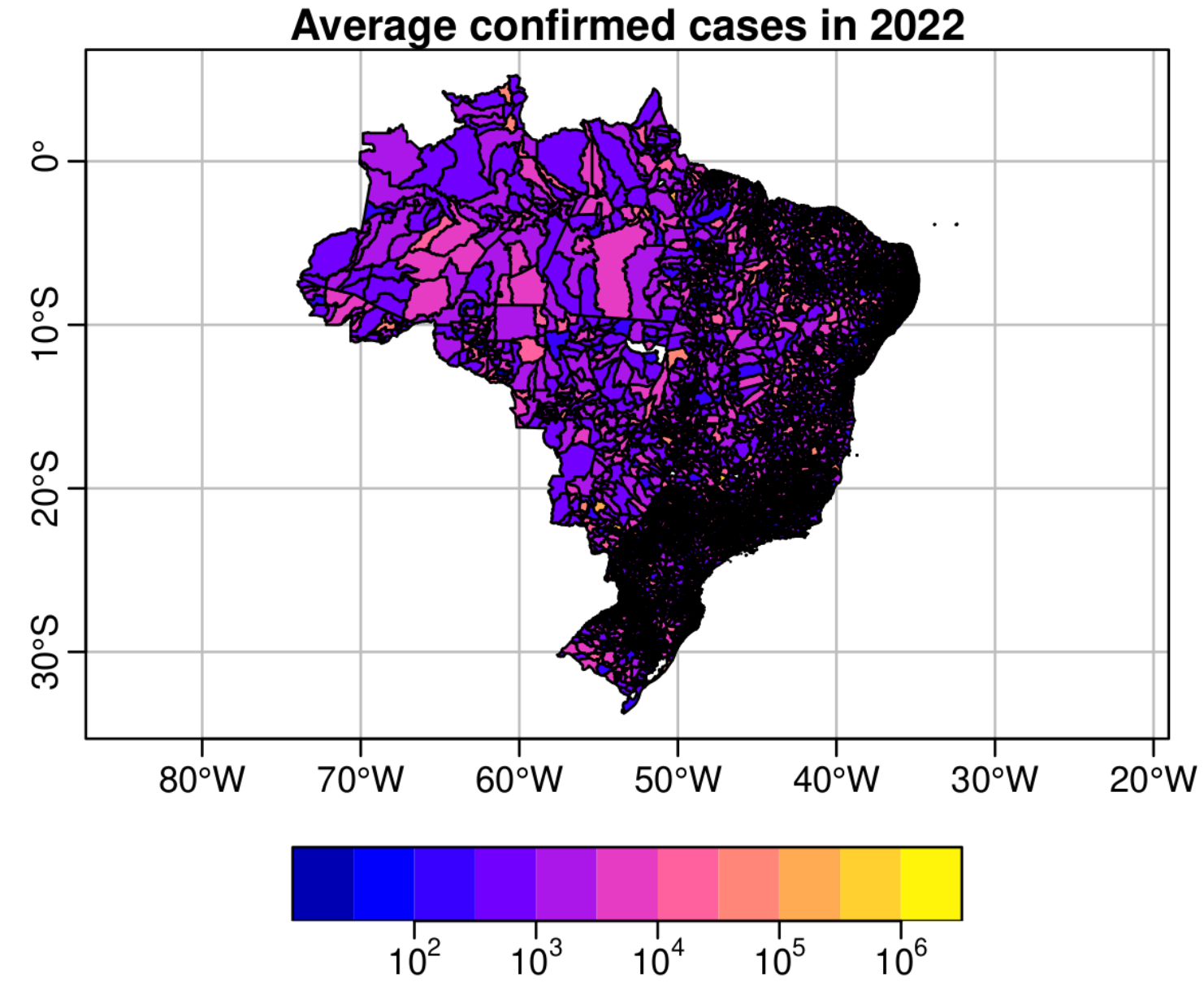}
\qquad
\includegraphics[width=8.6cm]{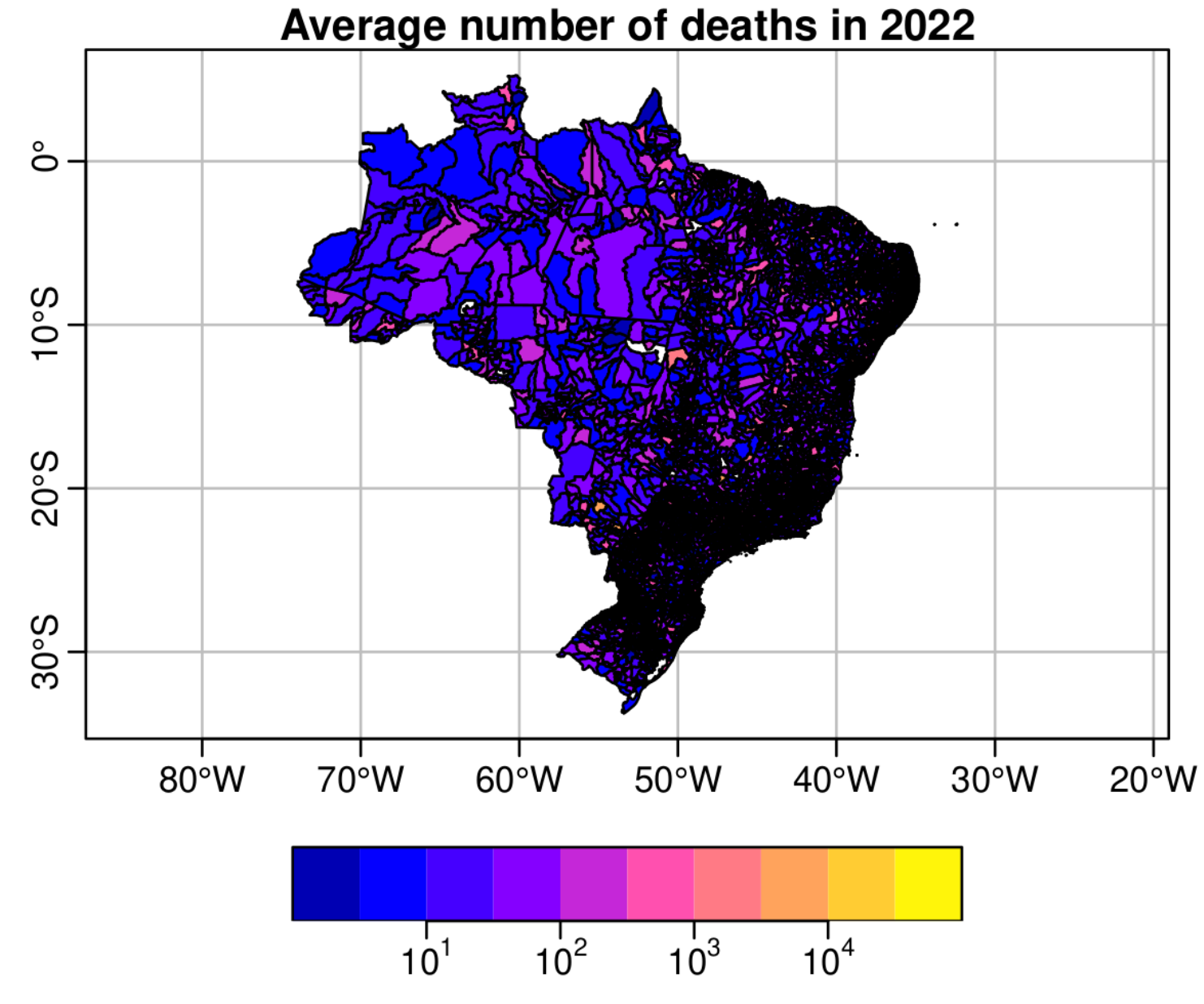}
\caption{\small{Choropleth maps depicting the average confirmed cases (left panels) and the average number of deaths (right panels) for Brazilian cities in the years 2021 (top panels) and 2022 (bottom panels)}.}\label{fig:Fig_2}
\end{figure}

Figure~\ref{fig:Fig_2} presents choropleth maps that vividly illustrate these spatial dependencies. The left panels depict the average confirmed cases, revealing clusters of cities with high case counts, particularly in urbanized regions. In contrast, the right panels show the average number of deaths, highlighting areas with elevated mortality rates. Notably, there exists a strong spatial correlation; cities with high confirmed cases often correspond to higher death rates, suggesting that geographical proximity influences the spread and severity of COVID-19 outcomes. This interdependence underscores the importance of considering spatial factors in analyzing health data, as the outcomes in one city are likely affected by conditions in neighboring areas, further reinforcing the necessity for spatially aware modeling in understanding the dynamics of the pandemic across Brazil. Recent research has demonstrated a significant spatial correlation between COVID-19-related deaths and confirmed cases, suggesting that neighboring cities often share similar epidemiological characteristics \citep[see, e.g.,][]{Saffary2020, Almalki2022, Khedhiri2022}. This correlation can be attributed to the movement of individuals across urban boundaries for work, social visits, and other interactions, which enhances the potential for infection transmission among adjacent areas.

\begin{figure}[!htb]
\centering
\includegraphics[width=8.6cm]{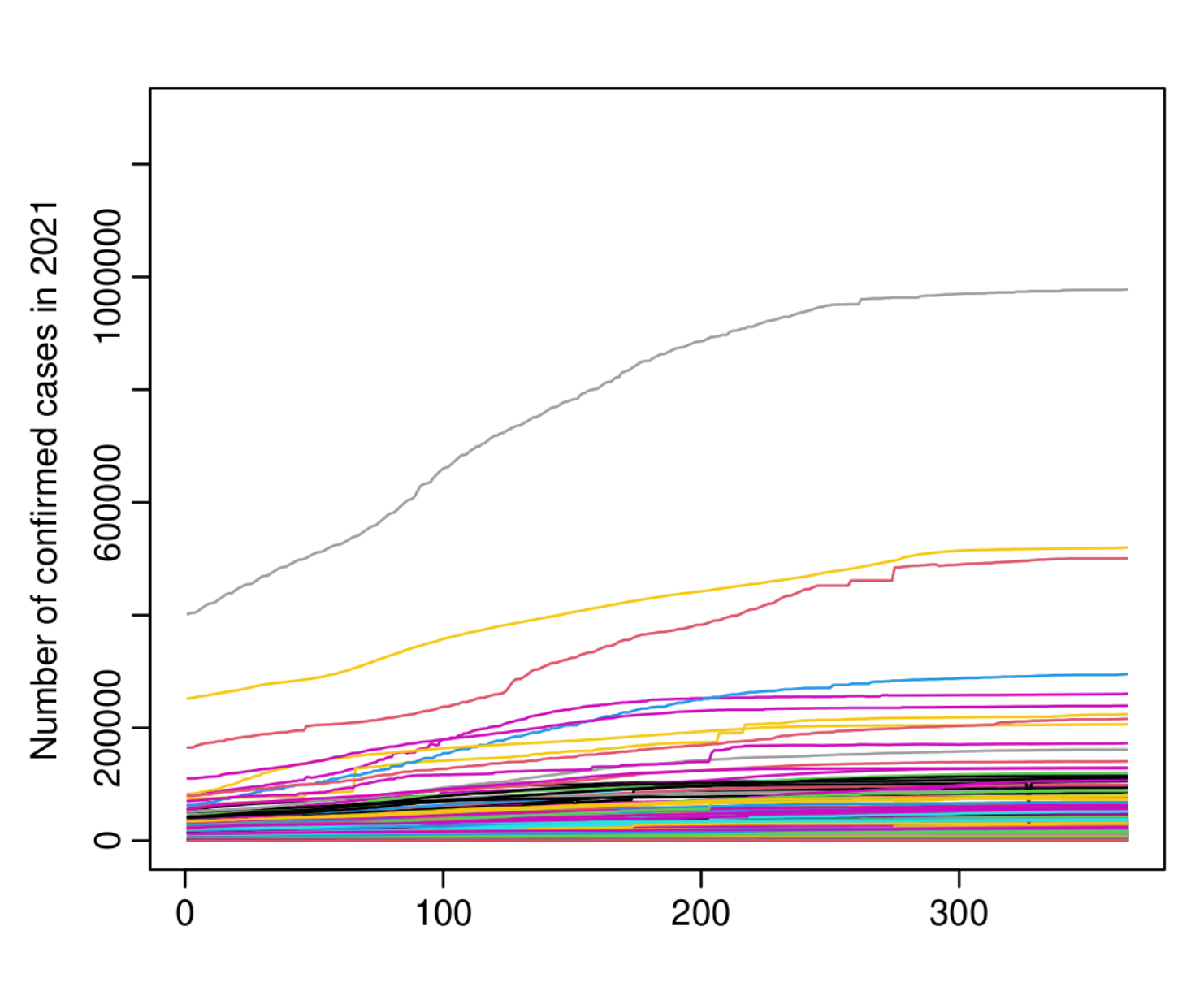}
\includegraphics[width=8.6cm]{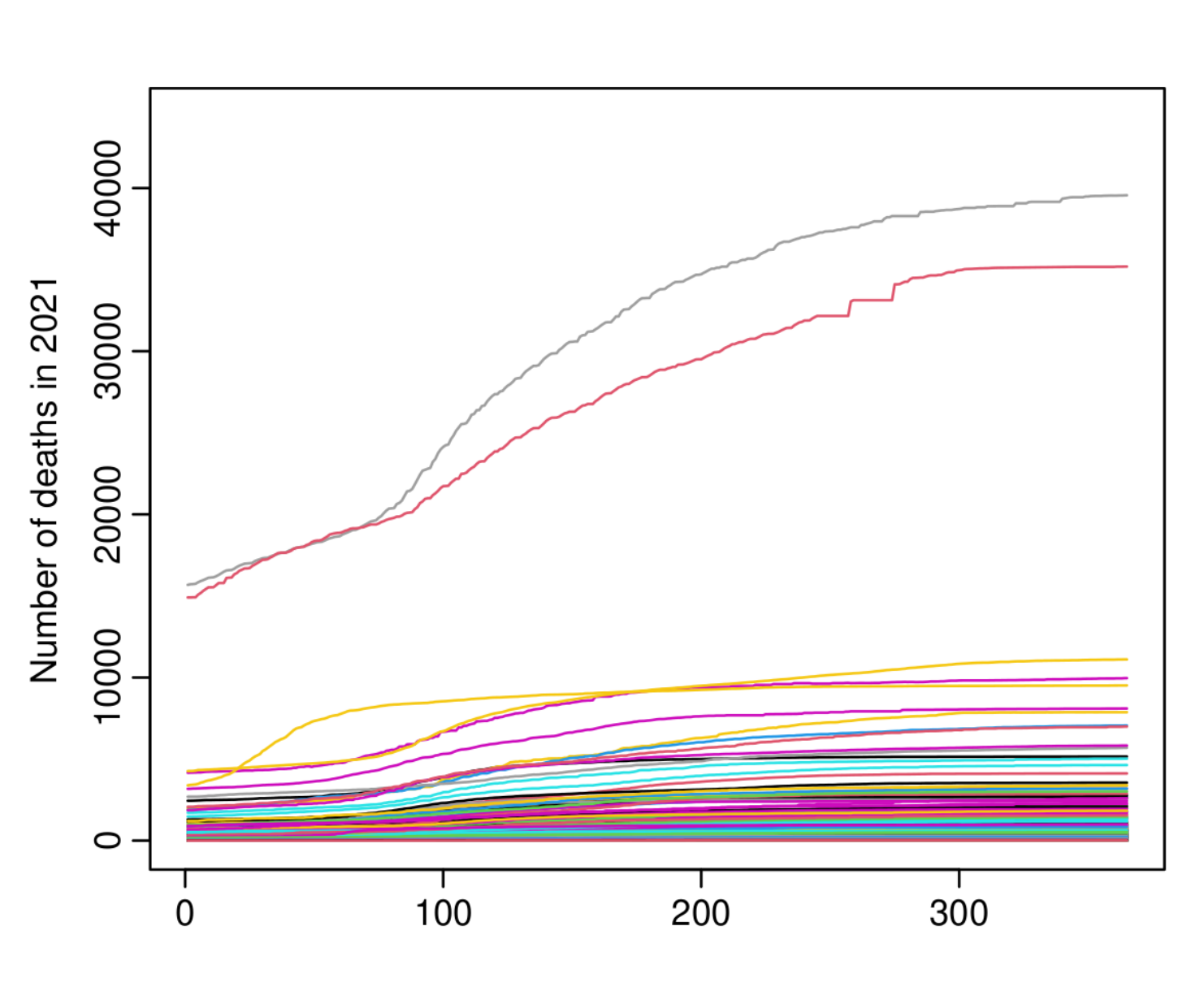}
\\ 
\includegraphics[width=8.6cm]{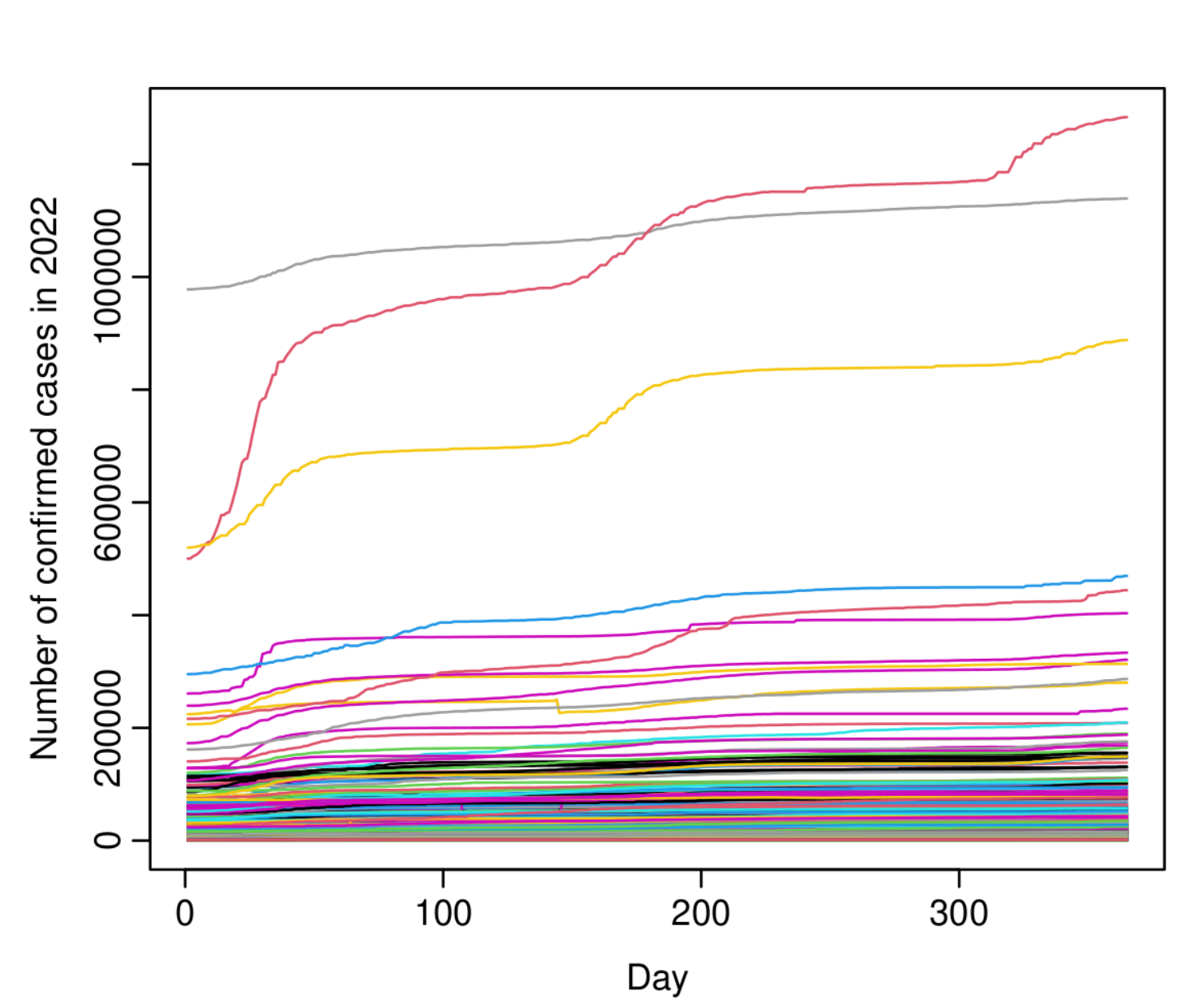}
\includegraphics[width=8.6cm]{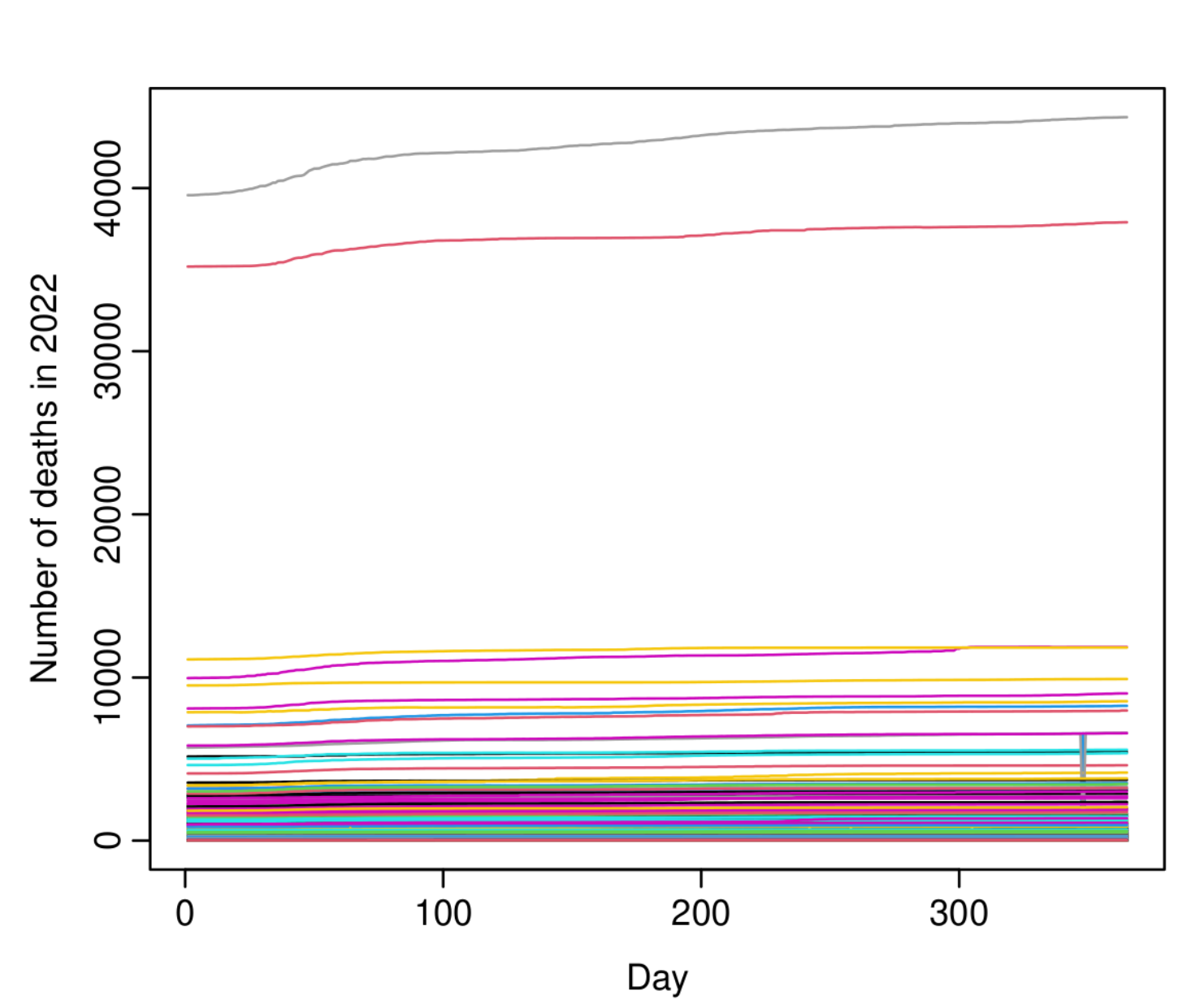}
\caption{\small{The graphical display of Brazil's COVID-19 data presents the confirmed cases in the left panels and the average number of deaths in the right panels for 2021 (top panels) and 2022 (bottom panels). Different colors represent different cities, with the observations of confirmed cases being functions of days, i.e., $1 \leq t \leq 365$}.}\label{fig:Fig_3}
\end{figure}

In this dataset, $\Y(s)$ denotes the daily number of deaths, while $\X(s)$ represents confirmed COVID-19 cases across Brazilian cities. Figure~\ref{fig:Fig_2} visually summarizes the daily deaths and confirmed cases for 2021 and 2022, revealing notable disparities among cities. As illustrated in Figures~\ref{fig:Fig_2} and~\ref{fig:Fig_3}, certain cities exhibit markedly elevated average numbers of deaths and confirmed cases, highlighting potential outliers in both the response and predictor variables. To mitigate the influence of these outliers on our analytical outcomes, we apply the natural logarithm transformation to both the response and predictor variables, ensuring a more robust and reliable analysis.

In our spatial correlation analysis of the Brazilian COVID-19 dataset, we compute the functional Moran’s I statistic to assess the degree of spatial autocorrelation. To begin, we construct the spatial weight matrix $\bm{W} = (w_{ii^{\prime}})_{n \times n}$ using the $K$-nearest neighbors (KNN) criterion, which is widely employed in practice. The distances between cities are calculated using the great circle distance between locations $i$ and $i^{\prime}$, derived from the Haversine formula. Specifically, this distance is given by $d_{ii^{\prime}} = Rc$, where $c = 2 \times \atantwo (\sqrt{a}, \sqrt{1-a})$ with $a = \sin^2(\Delta u / 2) + \cos(u_1) \cos(u_2) \sin^2(\Delta v / 2)$. Here, $u$ denotes the latitude, $v$ represents the longitude, and $R$ is the Earth’s mean radius (6,371 km). The KNN-based weight matrix is then defined as:
\begin{equation*} 
w_{ii^{\prime}} = \begin{cases} 
N_h(i)^{-1}, & \text{if}~ i^{\prime} \in N_h(i) \\
0, & \text{otherwise} 
\end{cases}
\end{equation*}
where $N_h(i)$ represents the set of $h$ nearest neighbors of location $i$, and $h$ is selected via a cross-validation approach.

The classical Moran’s I statistic, introduced by \cite{Ansellin1995}, measures the degree of spatial autocorrelation in a dataset, identifying whether similar values tend to cluster spatially and to what extent values at one location are influenced by neighboring locations. To assess spatial correlations in the functional response variable (daily number of deaths), we utilize the functional Moran’s I statistic, as outlined by \cite{Khoo2023}. First, this statistic is computed by representing the functional response variable through a basis expansion approach. Specifically, we employ a B-spline basis expansion, where $\bm{\varphi}(t) = [\varphi_1(t), \varphi_2(t), \ldots ]$ represents the set of B-spline basis functions and $\bm{Q} = [Q_1, Q_2, \ldots ]$ denotes the corresponding basis expansion coefficients. The functional Moran’s I statistic is then calculated as:
\begin{equation*}
I[\Y(t)] = \frac{\bm{\varphi}^\top(t) \bm{Q}^\top \bm{W} \bm{Q} \bm{\varphi}(t)}{\bm{\varphi}^\top(t) \bm{Q}^\top \bm{Q} \bm{\varphi}}.
\end{equation*}
This formulation allows us to quantify the spatial dependencies in the functional data, thereby providing insight into the clustering patterns of the response variable across neighboring locations.

\begin{figure}[!htb]
\centering
\includegraphics[width=8.8cm]{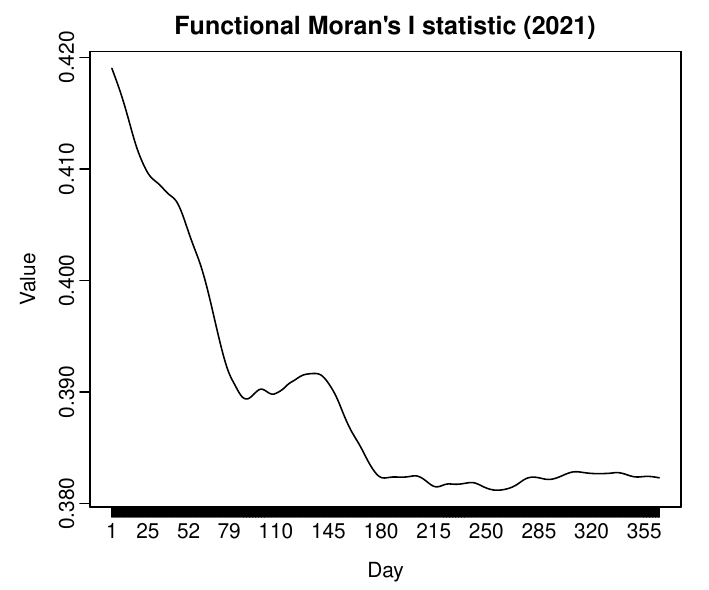}
\includegraphics[width=8.8cm]{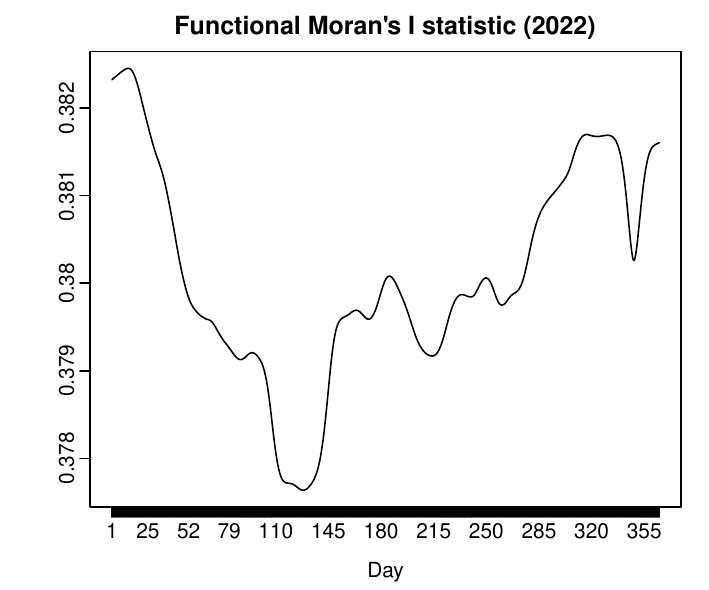}
\caption{\small{Functional Moran's I statistics for 2021 (left panel) and 2022 (right panel) were calculated using the KNN connection network.}}\label{fig:Fig_4}
\end{figure}

The functional Moran's I statistics presented in Figure~\ref{fig:Fig_4} illustrate the spatial correlation between Brazilian cities regarding COVID-19-related deaths for the years 2021 and 2022. In early 2021, the plot shows a positive spatial correlation, which systematically decreases until mid-year before stabilizing for the remainder of the year. One possible explanation for the decline in spatial correlation during the first half of 2021 could be government interventions aimed at restricting movement between cities to mitigate the spread of the virus. These restrictions likely disrupted the usual patterns of interaction and movement, which in turn weakened the spatial dependence between cities during this period. After mid-2021, the spatial correlation stabilizes, remaining steady through 2022. This could be attributed to the widespread implementation of COVID-19 vaccinations across the country, which helped bring the pandemic under control, leading to more homogeneous infection patterns. The relatively stable range of spatial correlation values in 2022, with the y-axis fluctuating between 0.378 and 0.382, suggests minimal changes in intercity relationships regarding confirmed cases and fatalities. This stability reflects the less dynamic nature of the pandemic in Brazil during 2022 compared to the fluctuations observed earlier.

We consider the following SFoFR model for analyzing the Brazilian COVID-19 dataset:
\begin{equation*}
\Y(t) = \bm{W} \int_{u=1}^{365} \Y(u) \rho(u,t) du + \int_{s=1}^{365} \X(s) \beta(s,t) ds + \epsilon(t).
\end{equation*}
We implement the proposed estimation procedure, alongside the classical FPLS and FPC approaches, to estimate the SFoFR model using the 2021 COVID-19 dataset. These fitted models are then employed to predict the daily number of deaths for 2022, using the number of confirmed cases as predictors. To evaluate the predictive performance of the models, we compute the MSE for in-sample predictions and the MSPE for out-of-sample predictions. Furthermore, to compare the in-sample and out-of-sample accuracy of the methods, we calculate the coefficient of determination ($R^2$) and the out-of-sample coefficient of determination ($R^2_{\text{new}}$) as follows:
\begin{align*}
R^2 &= 1 - \frac{\int_{t=1}^{365}\left\lbrace\Y(t) - \widehat{\Y}(t)\right\rbrace^2 dt}{\int_{t=1}^{365}\left\lbrace\Y(t) - \overline{\Y}(t)\right\rbrace^2 dt}, \\
R^2_{\text{new}} &= 1 - \frac{\int_{t=1}^{365}\left\lbrace\Y_{\text{new}}(t) - \widehat{\Y}_{\text{new}}(t)\right\rbrace^2 dt}{\int_{t=1}^{365}\left\lbrace\Y{\text{new}}(t) - \overline{\Y}{\text{new}}(t)\right\rbrace^2 dt},
\end{align*}
where $\widehat{\Y}(t)$ is the fitted daily number of deaths in 2021, $\Y_{\text{new}}(t)$ is the observed number of daily number of deaths in 2022, $\widehat{\Y}_{\text{new}}(t)$ is the predicted number of daily number of deaths in 2022, $\overline{\Y}(t)$ is the mean number of daily observed deaths in 2021, and $\overline{\Y}{\text{new}}(t)$ is the mean of number of daily observed deaths in 2022.
\begin{table}[!htb]
\centering
\tabcolsep 0.62in
\caption{\small{Computed MSE, MSPE, $R^2$, and $R^2_{\text{new}}$ values from the COVID-19 data.}}\label{tab:tab_3}
\begin{tabular}{@{}lcccc@{}} 
\toprule
{Method} & MSE & MSPE & $R^2$ & $R^2_{\text{new}}$ \\
\midrule
FPLS & 0.329 & 0.458 & 0.848 & 0.721 \\
FPC & 0.332 & 0.481 & 0.840 & 0.710 \\
SFoFR & 0.285 & 0.421 & 0.862 & 0.765 \\
\bottomrule
\end{tabular}
\end{table}

\begin{figure}[!htb]
\centering
\includegraphics[width=5.8cm]{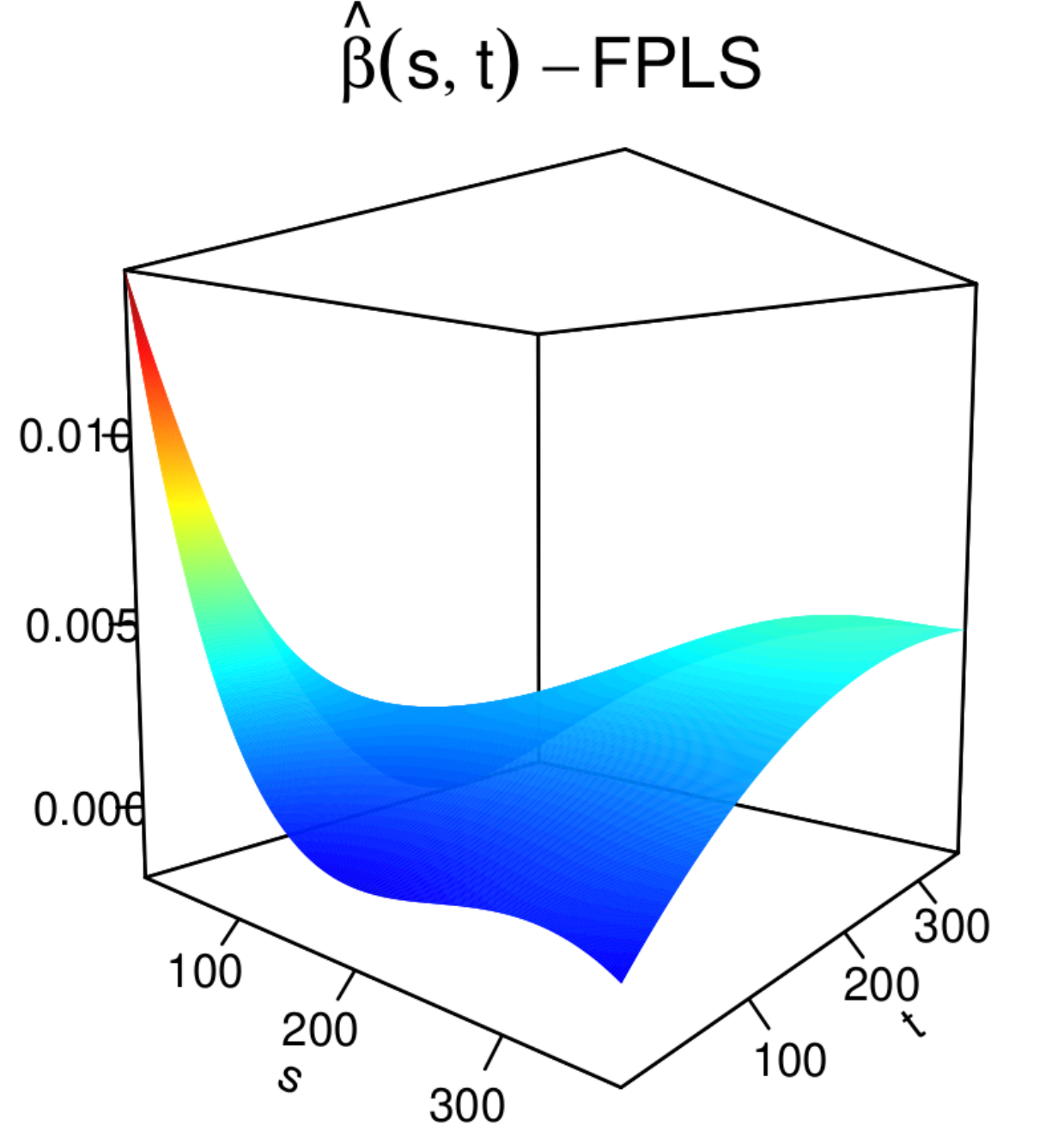}
\includegraphics[width=5.8cm]{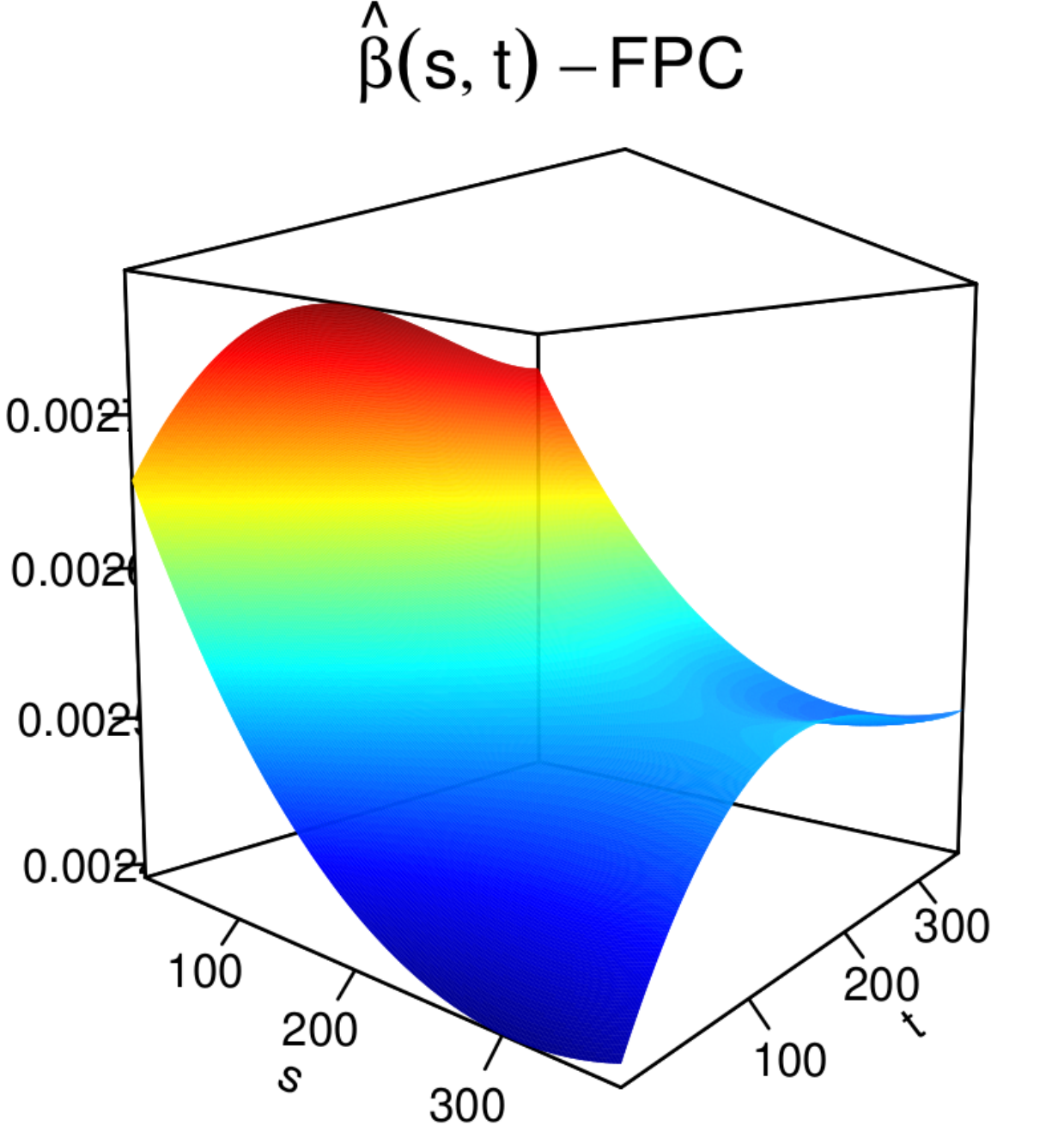}
\includegraphics[width=5.8cm]{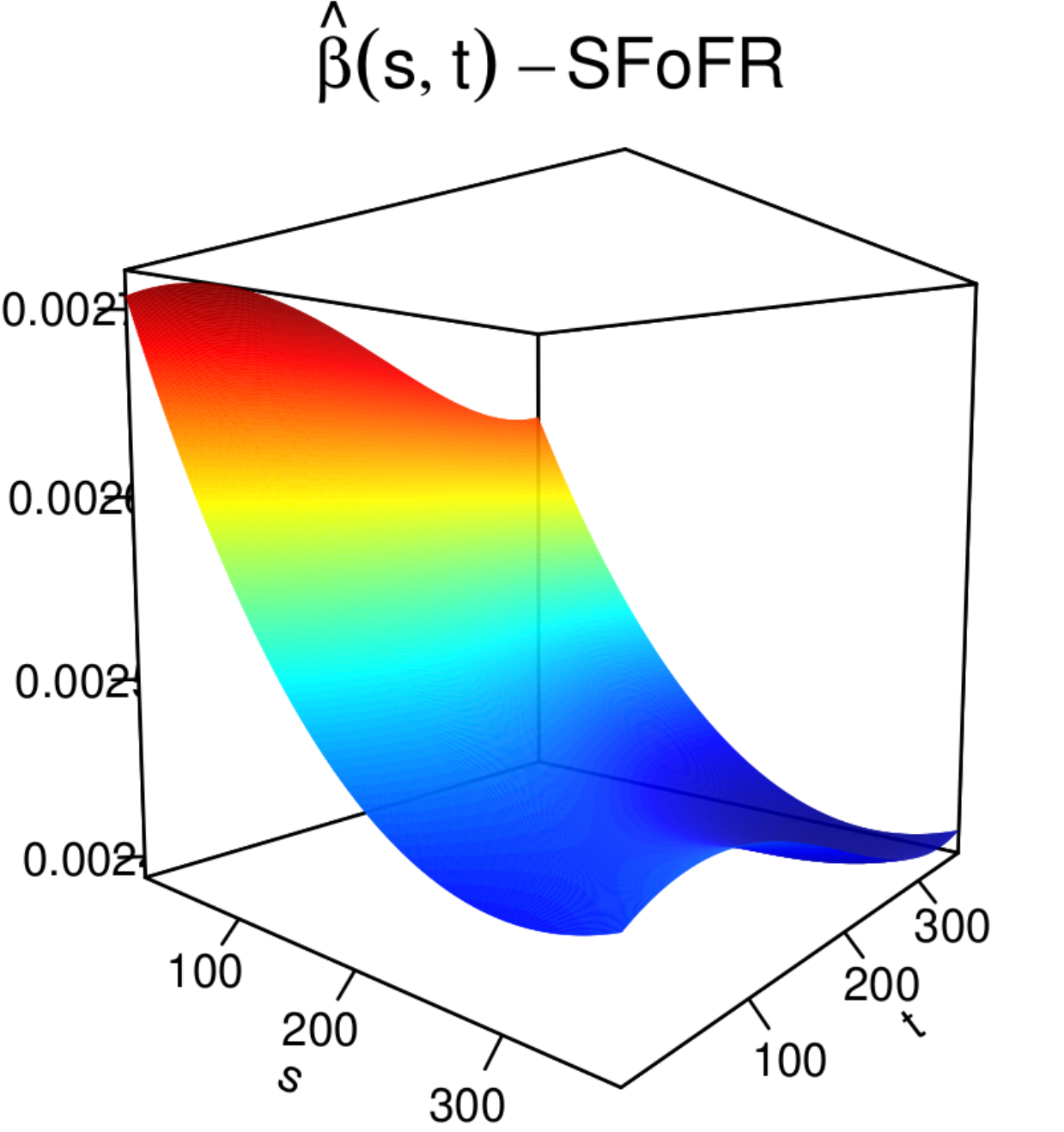}
\caption{\small{Estimated bivariate regression parameter functions for the Brazilian COVID-19 dataset obtained by the methods FPLS (left panel), FPC (middle panel), and SFoFR (right panel).}}\label{fig:Fig_5}
\end{figure}

The computed MSE, MSPE, $R^2$, and $R^2_{\text{new}}$ values are summarized in Table~\ref{tab:tab_3}. The results demonstrate that the proposed method achieves superior in-sample and out-of-sample predictive performance compared to the FPLS and FPC methods. Figure~\ref{fig:Fig_5} presents the estimated bivariate regression coefficient functions for all methods. These estimates reveal that the number of confirmed cases had a negative effect on the number of deaths during the first half of 2021 before stabilizing. All estimated bivariate regression coefficient functions align with the patterns identified by the functional Moran's I statistic, as shown in Figure~\ref{fig:Fig_4}.

\begin{figure}[!htb]
\centering
\includegraphics[width=8cm]{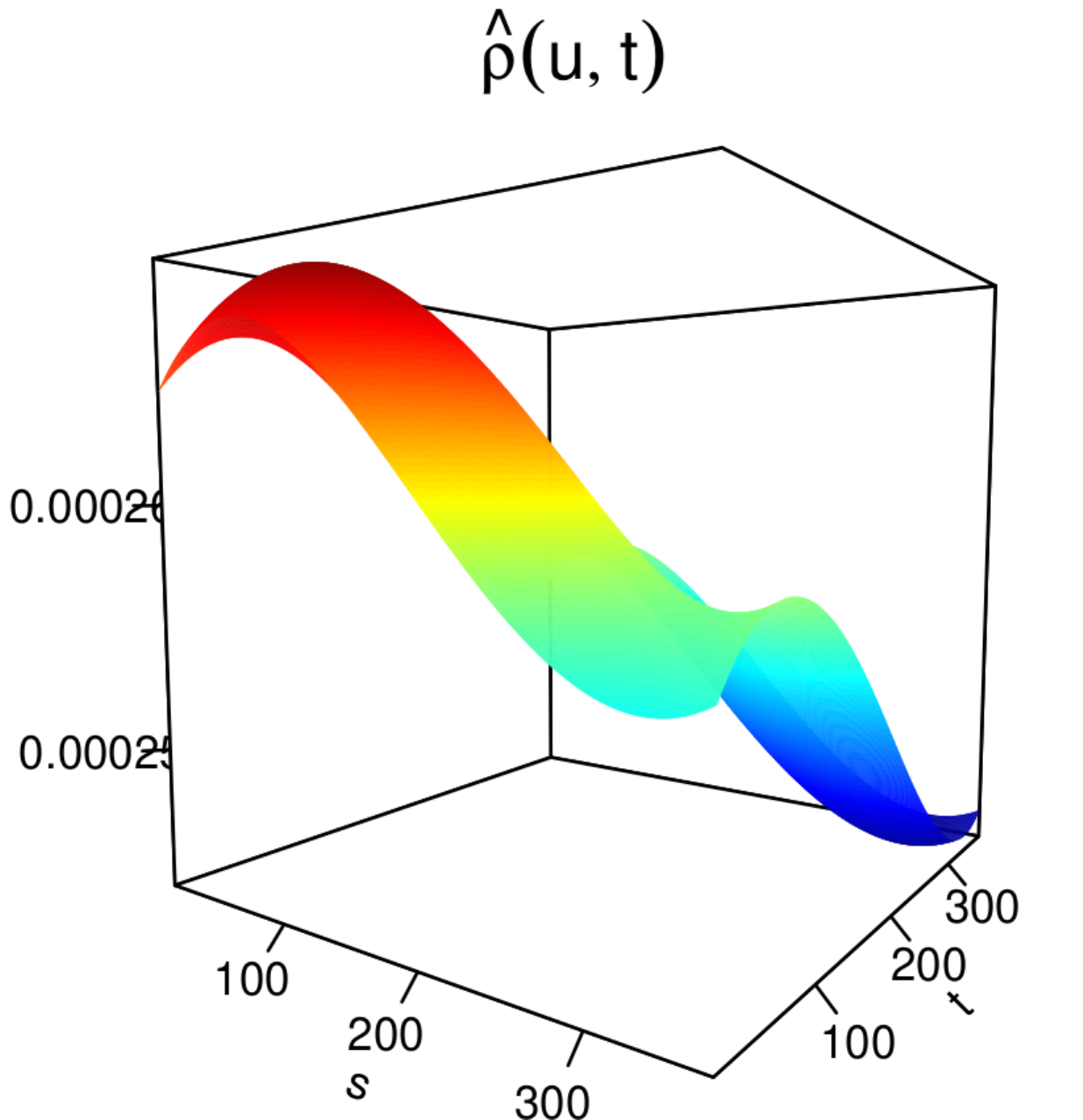}
\caption{\small{Estimated spatial autocorrelation parameter function for the Brazilian COVID-19 dataset by the proposed SFoFR.}}\label{fig:Fig_6}
\end{figure}

The estimated spatial autocorrelation parameter function, obtained using the proposed method, is depicted in Figure~\ref{fig:Fig_6}. This plot indicates that spatial correlation between Brazilian cities consistently decreases until mid-2021, after which it stabilizes, further supporting the findings from the functional Moran's I statistic. Overall, the proposed method effectively captures the spatial autocorrelation in the COVID-19 data, resulting in improved in-sample and out-of-sample predictions compared to the classical non-spatial methods (FPLS and FPC).

\section{Conclusion}\label{sec:6}

We introduced a novel SFoFR model to address complex spatial dependencies inherent in functional data. By combining spatial autoregressive modeling with FPC for functional predictors and responses, the SFoFR model offers a sophisticated framework for analyzing areal data where spatial interdependencies play a crucial role. Through a rigorous simulation study, we demonstrated that the proposed SFoFR model effectively captures spatial dependencies across different levels of spatial correlation, outperforming traditional FPC and FPLS methods in scenarios with moderate to strong spatial effects. The SFoFR model exhibited robust estimation and predictive accuracy, particularly in settings that require consideration of long-range and localized spatial effects.

Applying the SFoFR model to the Brazilian COVID-19 data provided valuable insights into the spatial dynamics of pandemic indicators across cities. The model's ability to account for intercity interactions in daily confirmed cases and death rates revealed that spatial dependency plays a significant role in COVID-19 outcomes, thus emphasizing the importance of spatially aware functional data analysis in epidemiological studies. This application demonstrated that incorporating spatial relationships into function-on-function regression frameworks can improve both interpretability and predictive accuracy, especially when modeling phenomena influenced by geographic and socio-economic factors.

The findings underscore the SFoFR model's potential as a powerful tool for diverse applications in fields where spatially dependent functional data are prevalent, including environmental sciences, public health, and economics. Computationally, its \Rlogo \ code is documented in the \texttt{[package anonymized for review]} package, available at (\texttt{anonimized url for review}). Future research directions may focus on extending the SFoFR framework to incorporate non-linear spatial effects, exploring its adaptability in other domains with complex spatial-temporal structures, and extending the proposed model to more complicated cases where the model allows for more than one functional predictor and scalar covariates in the model.

\section*{Acknowledgments}

This research was supported by the Scientific and Technological Research Council of Turkey (TUBITAK) (grant no. 124F096) and the Australian Research Council Future Fellowship (grant no. FT240100338). This research was also partly supported by the National Institute of Diabetes and Digestive and Kidney Diseases Award numbers 1R01DK132385-01 and R01DK136994-01A1. 

\newpage

\begin{center}
\large {\bf Supplementary material on ``Spatial function-on-function regression''}
\end{center}

\spacingset{1.45} 

In this supplement file, we provide technical details for the proof of Proposition 2.1 and proof of Theorem 3.1. Also, we present a detailed form of the covariance operator of the asymptotic Gaussian process in Theorem 3.1.

\begin{proof}[Proof of Proposition 2.1]
The proof is straightforward. For $\Y(t) \in (\mathcal{L}^p)^n[0,1]$, the operator $\mathcal{T}$ maps $(\mathcal{L}^p)^n[0,1]$ to itself. We first show that $\mathcal{T}$ is a contraction, that is, the operator norm of $\mathcal{T}$, which is defined as $\Vert \mathcal{T} \Vert = \sup_{\Vert \Y \Vert_{(\mathcal{L}^p)^n} \neq 0} \frac{\Vert \mathcal{T} \Y \Vert_{(\mathcal{L}^p)^n}}{\Vert \Y \Vert_{(\mathcal{L}^p)^n}}$ is less than 1. For $\Y(t) \in (\mathcal{L}^p)^n[0,1]$,
\begin{equation*}
\Vert \mathcal{T} \Y \Vert_{(\mathcal{L}^p)^n} = \left\lbrace \sum_{i=1}^n \Bigg\Vert \sum_{i^{\prime}=1}^n w_{i i^{\prime}} \int_0^1 \Y_{i^{\prime}}(u) \rho(u,t) du \Bigg\Vert_{\mathcal{L}^p}^p \right\rbrace^{1/p}.
\end{equation*}
Using Minkowski's inequality,
\begin{equation*}
\Bigg\Vert \sum_{i^{\prime}=1}^n w_{i i^{\prime}} \Y_{i^{\prime}}(u) \rho(u,t) du \Bigg\Vert_{\mathcal{L}^p} \leq \sum_{i^{\prime}=1}^n \vert w_{i i^{\prime}} \vert \Bigg\Vert \int_0^1 \Y_{i^{\prime}}(u) \rho(u,t) du \Bigg\Vert_{\mathcal{L}^p}.
\end{equation*}
By Jensen's inequality,
\begin{equation*}
\Bigg\Vert \int_0^1 \Y_{i^{\prime}}(u) \rho(u,t) du \Bigg\Vert_{\mathcal{L}^p} \leq \Vert \rho \Vert_{\infty} \Vert \Y_{i^{\prime}} \Vert_{\mathcal{L}^p}.
\end{equation*}
Hence,
\begin{equation*}
\Bigg\Vert \sum_{i^{\prime}=1}^n w_{i i^{\prime}} \Y_{i^{\prime}}(u) \rho(u,t) du \Bigg\Vert_{\mathcal{L}^p} \leq \Vert \rho \Vert_{\infty} \sum_{i^{\prime}=1}^n \vert w_{i i^{\prime}} \vert \Vert \Y_{i^{\prime}} \Vert_{\mathcal{L}^p}.
\end{equation*}
Summing over $i$ and using the definition of $\Vert \bm{W} \Vert_{\infty}$, we obtain $\Vert \mathcal{T} \Y \Vert_{(\mathcal{L}^p)^n} \leq \Vert \rho \Vert_{\infty} \Vert \bm{W} \Vert_{\infty} \Vert \Y \Vert_{(\mathcal{L}^p)^n}$. Given $\Vert \rho \Vert_{\infty} < \frac{1}{\Vert \bm{W} \Vert_{\infty}}$, it follows that $\Vert \mathcal{T} \Vert = \sup_{\Vert \Y \Vert_{(\mathcal{L}^p)p^n} \neq 0} \frac{\Vert \mathcal{T} \Y \Vert_{(\mathcal{L}^p)^n}}{\Vert \Y \Vert_{(\mathcal{L}^p)^n}} < 1$. Thus, $\mathcal{T}$ is a contraction.

Since $\mathcal{T}$ is a contraction on the complete metric space $(\mathcal{L}^p)^n[0,1]$, by the Banach Fixed-Point Theorem \citep[cf.][Chapter 1]{Kirk2001, Agarwal2018}, there exists a unique fixed point $\Y(t)$ such that
\begin{equation*}
\Y(t) = \mathcal{T} \Y(t) + \mathcal{G}(t),
\end{equation*}
where
\begin{equation*}
\mathcal{G}(t) = \int_0^1 \X(s) \beta(s,t) ds + \epsilon(t).
\end{equation*}
Since $\mathcal{T}$ is a contraction, the operator $(\mathbb{I}_d - \mathcal{T})^{-1}$ exists and can be expressed using the Neumann series expansion: $(\mathbb{I}_d - \mathcal{T})^{-1} = \sum_{k=0}^{\infty} \mathcal{T}^k$, which converges in the operator norm due to $\Vert \mathcal{T} \Vert < 1$. 

Given the existence of $(\mathbb{I}_d - \mathcal{T})^{-1}$, we can express the solution $\Y(t)$ as $\Y(t) = (\mathbb{I}_d - \mathcal{T})^{-1} \mathcal{G}(t)$. Substituting $\mathcal{G}(t)$, the following holds:
\begin{equation*}
\Y(t) = (\mathbb{I}_d - \mathcal{T})^{-1} \left[ \int_0^1 \X(s) \beta(s,t) ds + \epsilon(t) \right],
\end{equation*}
which completes the proof.
\end{proof}

Before the proof of Theorem 3.1, we first present the covariance operator $\Sigma_{\bm{\theta}}$ as follows:
\begin{equation*}
\bm{\Sigma}_{\bm{\theta}} =
\begin{pmatrix}
\bm{\Sigma}_{\rho}(u, t; u', t') & \bm{\Sigma}_{\rho\widetilde{\beta}}(u, t; s, t') \\
\bm{\Sigma}_{\widetilde{\beta}\rho}(s, t; u', t') & \bm{\Sigma}_{\widetilde{\beta}}(s, t; s', t')
\end{pmatrix},
\end{equation*}
where the components are given by 
\begin{align*}
\bm{\Sigma}_{\rho}(u, t; u', t') &= \phi^\top(u) \bm{\Sigma}_{0\rho} \phi(u') \phi^\top(t) \phi(t'), \quad \bm{\Sigma}_{\widetilde{\beta}}(s, t; s', t') = \psi^\top(s) \bm{\Sigma}_{0\beta} \psi(s') \phi^\top(t) \phi(t'), \\
\bm{\Sigma}_{\rho\widetilde{\beta}}(u, t; s, t') &= \phi^\top(u) \bm{\Sigma}_{0\rho\widetilde{\beta}} \psi(s) \phi^\top(t) \phi(t'), \quad \bm{\Sigma}_{\widetilde{\beta}\rho}(s, t; u', t') = \psi^\top(s) \bm{\Sigma}_{0\widetilde{\beta}\rho} \phi(u') \phi^\top(t) \phi(t'),
\end{align*}
and the finite-dimensional covariance matrices $\bm{\Sigma}_{0\rho}$, $\bm{\Sigma}_{0\widetilde{\beta}}$, $\bm{\Sigma}_{0\rho\widetilde{\beta}}$, and $\bm{\Sigma}_{0\widetilde{\beta}\rho}$ are defined as:
\begin{align*}
\bm{\Sigma}_{0\rho} &= (\bm{\Sigma}_{2\rho})^{-1} \bm{\Sigma}_{1\rho} (\bm{\Sigma}_{2\rho})^{-1}, \quad \bm{\Sigma}_{0\widetilde{\beta}} = (\bm{\Sigma}_{2\widetilde{\beta}})^{-1} \bm{\Sigma}_{1\widetilde{\beta}} (\bm{\Sigma}_{2\widetilde{\beta}})^{-1}, \\
\bm{\Sigma}_{0\rho\widetilde{\beta}} &= (\bm{\Sigma}_{2\rho})^{-1} \bm{\Sigma}_{1\rho\widetilde{\beta}} (\bm{\Sigma}_{2\widetilde{\beta}})^{-1}, \quad \bm{\Sigma}_{0\widetilde{\beta}\rho} = (\bm{\Sigma}_{2\widetilde{\beta}})^{-1} \bm{\Sigma}_{1\widetilde{\beta}\rho} (\bm{\Sigma}_{2\rho})^{-1},
\end{align*}
where $\bm{\Sigma}_{1\rho} = \big( \bm{\Sigma}_{1\rho}^{(t_1, t_2)} : 1 \leq t_1, t_2 \leq K_y^2 \big)$, $\bm{\Sigma}_{1\rho \widetilde{\beta}} = \big( \bm{\Sigma}_{1\rho \widetilde{\beta}}^{(t_1)} : 1 \leq t_1 \leq K_y^2 \big)^\top \in \mathbb{R}^{K_y^2 \times (K_x K_y)}$, $\bm{\Sigma}_{2\rho} = \big( \bm{\Sigma}_{2\rho}^{(t_1, t_2)} : 1 \leq t_1, t_2 \leq K_y^2 \big) \in \mathbb{R}^{K_y^2 \times K_y^2}$, and $\bm{\Sigma}_{2\rho \widetilde{\beta}} = \big( \bm{\Sigma}_{2\rho \widetilde{\beta}}^{(t_1)} : 1 \leq t_1 \leq K_y^2 \big) \in \mathbb{R}^{K_y^2 \times (K_x K_y)}$. Recall that $\widetilde{\bm{e}}^* = (\bm{\Sigma}_e^{1/2})^{-1} \widetilde{\bm{e}}$, and thus, we have $\mathrm{Cov}(\widetilde{\bm{e}}^*) = \mathbb{I}_{nK_y}$. Define $\bm{M}_1 = \bm{m} \widetilde{\bm{S}}^\top (\bm{\Sigma}_e^{1/2} \otimes \mathbb{I}_n)$, $\bm{M}_{2,k_1k_2} = \big( \bm{m}_{\rho,k_1k_2} \widetilde{\bm{S}}^\top + \bm{m} \widetilde{\bm{S}}_{\rho, k_1 k_2}^\top + \bm{m} \widetilde{\bm{S}}^\top \bm{S}_{\rho,k_1k_2} \bm{S}^{-1} \big) (\bm{\Sigma}_e^{1/2} \otimes \mathbb{I}_n)$, $\bm{M}_{3,k_1k_2} = \bm{m} \widetilde{\bm{S}}^\top \bm{S}_{\rho,k_1k_2} \bm{S}^{-1}$, and $\bm{J}_{1,k_1k_2} = \bm{M}_{3,k_1k_2} (\mathbb{I}_{K_y} \otimes \bm{X}^\top) \widetilde{\beta}$. We then have $\bm{F} = \bm{M}_1 \widetilde{\bm{e}}^*$, $\bm{F}_{\rho,k_1k_2} = \bm{M}_{2,k_1k_2} \widetilde{\bm{e}}^* + \bm{J}_{1,k_1k_2}$, and $\bm{F}_{\widetilde{\beta}} = -\bm{m} \widetilde{\bm{S}}^\top (\mathbb{I}_{K_y} \otimes \bm{X}^\top)$. Additionally, define $\bm{M}_{k_1k_2} = \bm{M}_1^\top \bm{M}_{2,k_1k_2}$, $\bm{J}_{k_1k_2} = \bm{M}_1^\top \bm{J}_{1,k_1k_2}$, and $\bm{H} = 2 \bm{F}_{\widetilde{\beta}}^\top \bm{M}_1$. Then, it can be verified that $Q_{\rho,k_1k_2} = 2 (\widetilde{\bm{e}}^*)^\top \bm{M}_{k_1k_2} \widetilde{\bm{e}}^* + 2 (\widetilde{\bm{e}}^*)^\top \bm{J}_{k_1k_2}$ and $Q_{\widetilde{\beta}} = \bm{H} \widetilde{\bm{e}}^*$. The following covariance matrices are then defined:
\begin{align*}
\bm{\Sigma}_{1\rho}^{(t_1, t_2)} &= \lim_{n \to \infty} \frac{1}{n} \big\{ 
4 \mathrm{tr}(\bm{M}_{k_1k_2} \bm{M}_{j_1j_2}^\top) + 
4 \mathrm{tr}(\bm{M}_{k_1k_2} \bm{M}_{j_1j_2}) + 4 \mathrm{tr}\big( \mathrm{diag}(\bm{M}_{k_1k_2}) \mathrm{diag}(\bm{M}_{j_1j_2}) \big) (\kappa_4 - 3) + 
4 \bm{J}_{k_1k_2}^\top \bm{J}_{j_1j_2}\big\}, \\
\bm{\Sigma}_{1\rho \widetilde{\beta}}^{(t_1)} &= 2 \lim_{n \to \infty} \frac{1}{n} \bm{H} \bm{J}_{k_1k_2}, \quad \bm{\Sigma}_{1 \widetilde{\beta}} = \lim_{n \to \infty} \frac{1}{n} \bm{H} \bm{H}^\top, \\
\bm{\Sigma}_{2\rho}^{(t_1, t_2)} &= 2 \lim_{n \to \infty} \frac{1}{n} \Big\{
\mathrm{tr}(\bm{M}_{2,k_1k_2}^\top \bm{M}_{2,j_1j_2}) + \bm{J}_{1,k_1k_2}^\top \bm{J}_{1,j_1j_2} \Big\}, \\
\bm{\Sigma}_{2\rho\widetilde{\beta}}^{(t_1)} &= 2 \lim_{n \to \infty} \frac{1}{n} \bm{F}_{\widetilde{\beta}}^\top \bm{J}_{1,k_1k_2}, \quad \bm{\Sigma}_{2\widetilde{\beta}} = 2 \lim_{n \to \infty} \frac{1}{n} \bm{F}_{\widetilde{\beta}}^\top \bm{F}_{\widetilde{\beta}},
\end{align*}
where $t_1 = (k_1 - 1)K_y + k_2$ and $t_2 = (j_1 - 1)K_y + j_2$ for $1 \leq k_1, k_2, j_1, j_2 \leq K_y$.

To establish the $\sqrt{n}$-consistency and asymptotic normality of the estimators, we outline the requisite conditions.
\begin{itemize}
\item[$C_1$] The functional processes $\Y(t)$ and $\X(s)$ possess a finite-dimensional Karhunen-Lo\`{e}ve decompositions, represented as $\Y(t) \approx \sum_{k_1=1}^{K_y} y_{k_1} \phi_{k_1}(t)$ and $\X(s) \approx \sum_{k_2=1}^{K_x} x_{k_2} \psi_{k_2}(s)$, respectively. The SFPC eigenfunctions $\{\phi_{k_1}(t)\}_{k_1=1}^{K_y}$ and FPC eigenfunctions $\{\psi_{k_2}(s)\}_{k_2=1}^{K_x}$ are uniformly bounded and orthonormal over their respective domains $[0, 1]$. They converge uniformly to the true eigenfunctions as $n \to \infty$, i.e.,
\begin{equation*}
\sup_{t \in [0, 1]} \|\widehat{\phi}_{k_1}(t) - \phi_{k_1}(t)\| \to 0 \quad \text{and} \quad \sup_{s \in [0, 1]} \|\widehat{\psi}_{k_2}(s) - \psi_{k_2}(s)\| \to 0 \quad \text{as } n \to \infty.
\end{equation*}
In addition, the scores $\bm{y}_{k_1} = \int \Y(t) \phi_{k_1}(t) dt$ and $\bm{x}_{k_2} = \int \X(s) \psi_{k_2}(s) ds$ satisfy:
\begin{equation*}
\mathbb{E}(\bm{y}_{k_1}^2) < \infty, \quad \mathbb{E}(\bm{x}_{k_2}^2) < \infty, \quad \text{and} \quad \frac{1}{n} \sum_{i=1}^n \bm{y}_{ik_1}^2 = \mathcal{O}(1), \quad \frac{1}{n} \sum_{i=1}^n \bm{x}_{ik_2}^2 = \mathcal{O}(1),
\end{equation*}
where $\bm{y}_{ik_1}$ and $\bm{x}_{ik_2}$ are the scores for the $i^\textsuperscript{th}$ spatial unit. Moreover, the estimated eigenvalues $\{\widehat{\lambda}_k\}$ of the covariance operators of $\Y(t)$ and $\X(s)$ satisfy $|\widehat{\lambda}_k - \lambda_k| \to 0 \quad \text{as } n \to \infty$ where $\lambda_k$ are the true eigenvalues. Furthermore, the eigenvalues decay at a rate such that:
\begin{equation*}
\sum_{k > K_y} \lambda_k = \mathcal{O}(K_y^{-p}) \quad \text{and} \quad \sum_{k > K_x} \lambda_k = \mathcal{O}(K_x^{-p}),
\end{equation*}
for some $p > 0$.    
\item[$C_2$] The error function $\epsilon(t)$ admits the Karhunen Lo\'eve decomposition $\epsilon(t) = \bm{e}^\top \bm{\phi}(t)$, where the random variable $\bm{e}$ is assumed to have mean $\bm{0}$ and $\mathrm{Cov}(\bm{e}) = \bm{\Sigma}_e \in \mathbb{R}^{K_y \times K_y}$.
\item[$C_3$] The regression parameter functions $\beta(s,t)$ and $\rho(u, t)$ lie in a linear subspace spanned by $\left\lbrace \phi_{k_1}(t) \right\rbrace_{k_1=1}^{K_y}$ and $\left\lbrace \psi_{k_2}(s) \right\rbrace_{k_2=1}^{K_x}$.
\item[$C_4$] The functional processes $\Y$ and $\X$ have finite fourth moments, i.e., $\mathbb{E} (\Vert \Y \Vert^4) < \infty$ and $\mathbb{E}(\Vert \X \Vert^4) < \infty$. 
\item[$C_5$] The symmetric matrix $\bm{W}^* = \bm{W} + \bm{W}^\top$ satisfies $\vert \lambda_1(\bm{W}^*) \vert = \mathcal{O}(\log~n)$.
\item[$C_6$] For any $\widetilde{\beta} \in \mathbb{R}^{K_y K_x}$, the covariates satisfy:
\begin{equation*}
\frac{1}{n} \Vert \bm{X}^{\top*} \widetilde{\beta} \Vert^2 = \mathcal{O}(1), \quad \text{as}~ n \rightarrow \infty,
\end{equation*}
where $\bm{X}^{\top*} = \mathbb{I}_{K_y} \otimes \bm{X}^{\top}$ and $\widetilde{\beta}$ is the vectorized regression coefficient matrix.
\item[$C_7$] Let $\bm{\Sigma}_e$ be decomposed as $\bm{\Sigma}_e = (\bm{\Sigma}_e^{1/2})^\top (\bm{\Sigma}_e^{1/2})$. The error vector $\widetilde{\bm{e}}^* = (\bm{\Sigma}_e^{1/2})^{-1} \widetilde{\bm{e}}$ satisfies $\mathbb{E}\{(\widetilde{e}_{ik}^*)^4\} = \kappa_4$, for $i = 1, \ldots, n$ and $k = 1, \ldots, K_y$, where $\kappa_4$ is a finite constant, and $\mathbb{E}(\widetilde{e}_{ik_1}^* \widetilde{e}_{ik_2}^* \widetilde{e}_{ik_1}^*) =~0$ for $k_1, k_2, k_3 = 1, \ldots, K_y$.
\item[$C_8$] Assume that the limit $\lim_{n \rightarrow \infty} n^{-1} [\bm{X}^{\ddagger} (\bm{X}^{\ddagger})^\top]$ exists and is nonsingular.
\item[$C_9$] Let $\widetilde{\bm{S}} = (\bm{\Omega}_e \otimes \mathbb{I}_n) \bm{S}$ and $\bm{F} = \bm{m} \widetilde{\bm{S}}^\top \big( \bm{S} \widetilde{Y} - (\bm{I}_{K_y} \otimes \bm{X}^\top) \widetilde{\beta} \big)$. The least squares objective function is then given by $Q(\bm{\rho}, \widetilde{\beta}, \bm{\Sigma}_e) = \bm{F}^\top \bm{F} $. Denote the first-order derivatives of $Q(\bm{\rho}, \widetilde{\beta}, \bm{\Sigma}_e)$ as $Q_{\rho,k_1k_2} = \partial Q / \partial \rho_{k_1k_2}$ and $Q_{\widetilde{\beta}} = \partial Q / \partial \bm{\widetilde{\beta}}$, and the second-order derivatives as $Q_{\rho,k_1k_2,j_1j_2} = \partial^2 Q / \partial \rho_{k_1k_2} \partial \rho_{j_1j_2}$, $Q_{\rho,k_1k_2,\widetilde{\beta}} = \partial^2 Q / \partial \rho_{k_1k_2} \partial \widetilde{\beta}$, and $Q_{\widetilde{\beta} \widetilde{\beta}} = \partial^2 Q / \partial \widetilde{\beta} \partial\widetilde{\beta}^\top$. Assume the following limits exist: 
\begin{align*}
\lim_{n \to \infty} \frac{1}{n} \mathrm{Cov}(Q_{\rho,k_1k_2} Q_{\rho,j_1j_2}) &= \bm{\Sigma}_{1\rho}^{(t_1, t_2)}, \quad \lim_{n \to \infty} \frac{1}{n} \mathrm{Cov}(Q_{\rho,k_1k_2}, Q_{\widetilde{\beta}}) = \bm{\Sigma}_{1\rho\beta}^{(t_1)} \in \mathbb{R}^{K_x K_y}, \\
\lim_{n \to \infty} \frac{1}{n} \mathrm{Cov}(Q_{\widetilde{\beta}}, Q_{\widetilde{\beta}}) &= \bm{\Sigma}_{1 \widetilde{\beta}} \in \mathbb{R}^{(K_x K_y) \times (K_x K_y)}, \quad \lim_{n \to \infty} \frac{1}{n} \mathbb{E}[Q_{\rho,k_1k_2,j_1j_2}] = \bm{\Sigma}_{2\rho}^{(t_1, t_2)}, \\
\lim_{n \to \infty} \frac{1}{n} \mathbb{E}[Q_{\rho,k_1k_2, \widetilde{\beta}}] &= \bm{\Sigma}_{2\rho \widetilde{\beta}}^{(t_1)} \in \mathbb{R}^{K_x K_y}, \quad \lim_{n \to \infty} \frac{1}{n} \mathbb{E}[Q_{\widetilde{\beta} \widetilde{\beta}}] = \bm{\Sigma}_{2\widetilde{\beta}} \in \mathbb{R}^{(K_x K_y) \times (K_x K_y)}.
\end{align*}
\end{itemize}

Conditions $C_1$-$C_3$ are fundamental to ensure that the infinite-dimensional SFoFR model can be effectively represented with a finite number of truncation constants denoted as $K_y$ and $K_x$. The satisfaction of Conditions $C_1$-$C_3$ is contingent upon the validity of $C_4$, a common condition in the asymptotic properties of FPC/SFPC. Conditions $C_6$-$C_9$ are required to show the $\sqrt{n}$-consistency and asymptotic normality of the estimators of the model constructed in the finite-dimensional space using FPC and SFPC coefficients.

\begin{proof}[Proof of Theorem 3.1]
Let $\mathbb{P}_{\Y}$ and $\mathbb{P}_{\X}$ respectively denote the image measures of $\Y$ and $\X$, i.e., $\mathbb{P}_{\Y}(U) = P(\Y \in U)$ and $\mathbb{P}_{\X}(V) = P(\X \in V)$  for Borel sets $U$ and $V$. The CDFs of $\Y$ and $\X$ are defined by:
\begin{align*}
F_{\Y}(d_1, \ldots, d_{K_y}) &:= \mathbb{P}_{\Y}( y_1 \leq d_1, \ldots, y_{K_y} \leq d_{K_y}), \\
F_{\X}(b_1, \ldots, b_{K_x}) &:= \mathbb{P}_{\X}( x_1 \leq b_1, \ldots, x_{K_x} \leq b_{K_x}).
\end{align*}
Denote by $F_{\epsilon}$ the distribution function of $\epsilon(t)$ similar to $F_{\Y}$. Then, the functionals of the least-squares estimators $\widehat{\beta}(s,t)$ and $\widehat{\rho}_{\tau}(u,t)$ are defined as follows:
\begin{align*}
\widehat{\beta}(F_{\X}, F_{\Y}, F_{\epsilon})(s,t) &= \sum_{k_2 = 1}^{K_{x}} \sum_{k_1 = 1}^{K_{y}} \beta_{k_2 k_1}(F_{\X}, F_{\Y}, F_{\epsilon}) \psi_{k_2}(F_{\X})(s) \phi_{k_1}(F_{\Y})(t), \\
\widehat{\rho}(F_{\Y}, F_{\Y}, F_{\epsilon})(u,t) &= \sum_{k^{\prime}_1 = 1}^{K_{\Y}} \sum_{k_1 = 1}^{K_{\Y}} \rho_{k^{\prime}_1 k_1}(F_{\Y}, F_{\Y}, F_{\epsilon}) \phi_{k^{\prime}_1}(F_{\Y})(u) \phi_{k_1} (F_{\Y})(t).
\end{align*}

By conditions $C_1$--$C_3$, we have $\Y(t) = \bm{Y}^\top \bm{\phi}(F_{\Y})(t)$, $\X(s) = \bm{X}^\top \bm{\psi}(F_{\X})(s)$, and $\epsilon(t) = e^\top \bm{\phi}(F_{\Y})(t)$. Then, by orthonormalities of $\bm{\phi}(F_{\Y})(t)$ and $\bm{\psi}(F_{\X})(s)$, we have
\begin{align*}
\bm{Y}^\top \bm{\phi}(F_{\Y})(t) =& \bm{W} \bm{Y}^\top \bm{\phi}(F_{\Y})(u) \bm{\phi}^\top(F_{\Y})(u) \bm{\rho}(F_{\Y}, F_{\Y}, F_{\epsilon}) \bm{\phi}(F_{\Y})(t) \\
&+ \bm{X}^\top \bm{\psi}(F_{\X})(s) \bm{\psi}^\top (F_{\X})(s) \bm{\beta}(F_{\X}, F_{\Y}, F_{\epsilon}) \bm{\phi}(F_{\Y})(t) + \bm{e}^\top \bm{\phi}(F_{\Y})(t), \\
\bm{Y}^{\top} =& \bm{W} \bm{Y}^\top \bm{\rho}(F_{\Y}, F_{\Y}, F_{\epsilon}) + \bm{X}^\top \bm{\beta}(F_{\X}, F_{\Y}, F_{\epsilon}) + \bm{e}^\top.
\end{align*}

The results given above show that, by conditions $C_1$-$C_3$, the infinite-dimensional SFoFR model is represented in the finite-dimensional space of FPC and SFP coefficients (i.e., MSAR model) as follows:
\begin{equation*}
\bm{Y}^{\top} = \bm{W} \bm{Y}^\top \bm{\rho} + \bm{X}^\top \bm{\beta} + \bm{e}^{\top}.
\end{equation*}
Let us now consider the vectorized representation of the MSAR model as follows:
\begin{equation*}
\widetilde{Y} = (\bm{\rho}^\top \otimes \bm{W}) \widetilde{Y} + \bm{X}^{\top*} \widetilde{\beta} + \widetilde{e},
\end{equation*}
where $\widetilde{Y} = \text{vec}(\bm{Y}^\top) = (Y_1, \ldots, Y_{K_y})^\top \in \mathbb{R}^{n K_y}$, $\bm{X}^{\top*} = \mathbb{I}_{K_y} \otimes \bm{X}^{\top}$, $\widetilde{e} = \text{vec}(\bm{e}^\top) = (e_1, \ldots, e_{K_y})^\top \in \mathbb{R}^{n K_y}$, and $\widetilde{\beta} = \text{vec}(\bm{\beta}) \in \mathbb{R}^{K_y K_x}$. Let $\widehat{\bm{\Theta}}^* = (\widehat{\widetilde{\bm{\rho}}}^\top, \widehat{\widetilde{\bm{\beta}}}^\top)$ denote the least-squares estimates of $\bm{\Theta}^* = (\widetilde{\bm{\rho}}^\top, \widetilde{\bm{\beta}}^\top)$. Next, we will prove the $\sqrt{n}$-consistency and asymptotic normality of $\widehat{\bm{\Theta}}^{*}$.

First, similar to \cite{Zhu2020}, for any $\varepsilon > 0$, we aim to show that
\begin{equation*}
\lim_{n \rightarrow \infty} \text{Pr} \left\lbrace \inf_{\Vert \bm{b} \Vert = \mathcal{B}} Q(\bm{\Theta}^* + n^{-1/2} \bm{b}) > Q(\bm{\Theta}^*) \right\rbrace \geq 1 - \varepsilon,
\end{equation*}
where $0 < \mathcal{B} < \infty$. Using the second-order Taylor expansion of $\bm{\Theta}^*$, we have
\begin{equation}\label{eq:tylr}
\inf_{\Vert \bm{b} \Vert = \mathcal{B}} \left\lbrace Q(\bm{\Theta}^* + n^{-1/2} \bm{b}) - Q(\bm{\Theta}^*) \right\rbrace = \mathcal{B} n^{-1/2} \frac{\partial Q(\bm{\Theta}^*)}{\partial \bm{\Theta}^*} \bm{b} + 2^{-1} \mathcal{B}^2 n^{-1} \bm{b}^\top \frac{\partial^2 Q(\bm{\Theta}^*)}{\partial \bm{\Theta}^* (\bm{\Theta}^*)^\top} \bm{b} + o_p(1)
\end{equation}
By Lemma~2 of \cite{Zhu2020}, $\frac{\partial Q(\bm{\Theta}^*)}{\partial \bm{\Theta}^*}$ has finite variance, and, thus, the linear term $n^{-1/2} \frac{\partial Q(\bm{\Theta}^*)}{\partial \bm{\Theta}^*} \bm{b}$ is $\mathcal{O}_p(1)$. The quadratic term, $n^{-1} \bm{b}^\top \frac{\partial^2 Q(\bm{\Theta}^*)}{\partial \bm{\Theta}^* (\bm{\Theta}^*)^\top} \bm{b}$ depends on the eigenvalues of $n^{-1} \frac{\partial^2 Q(\bm{\Theta}^*)}{\partial \bm{\Theta}^* (\bm{\Theta}^*)^\top}$. By condition~$C_9$ and Lemma 6 of \cite{Zhu2020}, we have $n^{-1} \frac{\partial^2 Q(\bm{\Theta}^*)}{\partial \bm{\Theta}^* (\bm{\Theta}^*)^\top} \overset{p}{\to} \bm{\Sigma}_L^{(2)}$, where
\begin{equation*}
\bm{\Sigma}_L^{(2)} =
\begin{pmatrix}
\bm{\Sigma}_{2\rho} & \bm{\Sigma}_{2\rho \widetilde{\bm{\beta}}} \\
\bm{\Sigma}_{2\rho \widetilde{\bm{\beta}}}^\top & \bm{\Sigma}_{2 \widetilde{\bm{\beta}}}
\end{pmatrix},
\end{equation*}
and $\lambda_{\text{min}} \lbrace n^{-1} \frac{\partial^2 Q(\bm{\Theta}^*)}{\partial \bm{\Theta}^* (\bm{\Theta}^*)^\top} \rbrace \overset{p}{\to} \lambda_{\text{min}} (\bm{\Sigma}_L^{(2)}) > 0$, asymptotically. Thus, the quadratic term in~\eqref{eq:tylr} is asymptotically positive for sufficiently large $\mathcal{B}$. In conclusion, because the coefficient of the linear term in~\eqref{eq:tylr} is bounded and the coefficient of the quadratic term is asymptotically positive, the Taylor expansion ensures that a local minimizer $\widehat{\bm{\Theta}}^*$ exists in the ball $\lbrace \bm{\Theta}^* + n^{-1/2} \bm{b} \mathcal{B}: \Vert \bm{b} \Vert \leq 1 \rbrace$. This result indicates that $\widehat{\bm{\Theta}}^*$ is $\sqrt{n}$-consistent.

Next, applying the Taylor expansion around $\bm{\Theta}^*$, we have:
\begin{equation*}
n^{-1} (\bm{\Theta}^* - \bm{\Theta}) = \left\lbrace n^{-1} \frac{\partial^2 Q(\bm{\Theta}^{**})}{\partial \bm{\Theta}^{**} (\bm{\Theta}^{**})^\top} \right\rbrace^{-1} \left\lbrace n^{-1/2} \frac{\partial Q(\bm{\Theta}^*)}{\partial \bm{\Theta}^*} \right\rbrace,
\end{equation*}
where $\bm{\Theta}^{**}$ is between $\bm{\Theta}^*$ and $\bm{\widehat{\Theta}}^*$. By the conclusion $n^{-1} \frac{\partial^2 Q(\bm{\Theta}^*)}{\partial \bm{\Theta}^* (\bm{\Theta}^*)^\top} \overset{p}{\to} \bm{\Sigma}_L^{(2)}$ and Lemma 6 in \cite{Zhu2020}, we have $n^{-1} \frac{\partial^2 Q(\bm{\Theta}^{**})}{\partial \bm{\Theta}^{**} (\bm{\Theta}^{**})^\top} \overset{p}{\to} \bm{\Sigma}_L^{(2)}$. By Lemmas 3-4 in \cite{Zhu2020} and condition~$C_9$, we also have $n^{-1/2} \frac{\partial Q(\bm{\Theta}^*)}{\partial \bm{\Theta}^*} \xrightarrow{d} \mathcal{N}(\bm{0}, \bm{\Sigma}_L^{(1)})$ as $n \rightarrow \infty$, where $\bm{\Sigma}_L^{(1)}$ is the block covariance matrix as follows:
\begin{equation*}
\bm{\Sigma}_L^{(1)} =
\begin{pmatrix}
\bm{\Sigma}_{1\rho} & \bm{\Sigma}_{1\rho\beta} \\
\bm{\Sigma}_{1\rho\beta}^\top & \bm{\Sigma}_{1\beta}
\end{pmatrix}.
\end{equation*}
Combining the two results, we have:
\begin{equation*}
\sqrt{n} (\widehat{\bm{\Theta}}^* - \bm{\Theta}^*) \xrightarrow{d} \mathcal{N}\big(\bm{0}, \bm{\Sigma}_L^{(2)^{-1}} \bm{\Sigma}_L^{(1)} \bm{\Sigma}_L^{(2)^{-1}}\big).
\end{equation*}

In conclusion, by conditions $C_1$-$C_9$, the linear mappings from the finite-dimensional coefficients to the functional space preserve asymptotic normality. Specifically:
\begin{align*}
\sqrt{n} \big( \widehat{\rho}(u, t) - \rho(u, t) \big) &\xrightarrow{d} \mathcal{GP}\big(0, \bm{\Sigma}_{\rho}(u, t; u', t')\big), \\
\sqrt{n} \big( \widehat{\beta}(s, t) - \beta(s, t) \big) &\xrightarrow{d} \mathcal{GP}\big(0, \bm{\Sigma}_{\widetilde{\beta}}(s, t; s', t')\big),
\end{align*}
which completes the proof.
\end{proof}

\bibliographystyle{agsm}
\bibliography{sfof.bib}

\end{document}